\documentstyle[12pt]{article}

\title{Symmetric random matrices and the Pfaff lattice}

\author{M. Adler\thanks{Department of Mathematics,
Brandeis University, Waltham, Mass 02454, USA. E-mail:
adler@math.brandeis.edu. The support of a National Science
Foundation grant \# DMS-9503246 is gratefully
acknowledged.}~~~~~~
P. van Moerbeke\thanks{Department of Mathematics, Universit\'e de
Louvain, 1348 Louvain-la-Neuve, Belgium and Brandeis University,
Waltham, Mass 02454, USA. E-mail: vanmoerbeke@geom.ucl.ac.be and
@math.brandeis.edu. The  support of a National Science Foundation
grant \# DMS-9503246, a Nato, a FNRS and a Francqui Foundation
grant is gratefully acknowledged.}}

\date{August 24, 1998}

\newcommand{\MAT}[1]{\left(\begin{array}{*#1c}}
\newcommand{\mat}{\end{array}\right)}
\newcommand{\qed}
{%
\mbox{}%
\nolinebreak%
\hfill%
\rule{2mm}{2mm}%
\medbreak%
\par%
}

\newcommand{\sumbis}[2]%
{%

\begin{array}[t]{c}
\sum\\
{\scriptstyle #1}\\
{\scriptstyle #2}
\end{array}

}

\newcommand{\rg}{\rightarrow}

\newcommand{\TT}{\tilde\tau}
\newcommand{\DR}{{\cal D}}
\newcommand{\LR}{{\cal L}}

\newcommand{\BC}{{\Bbb C}}

\newcommand{\BY}{{\Bbb Y}}
\newcommand{\BZ}{{\Bbb Z}}
\newcommand{\Sg}{\Sigma}
\newcommand{\iy}{\infty}
\newcommand{\pl}{\partial}
\newcommand{\al}{\alpha}
\newcommand{\proof}{\underline{\sl Proof}: }
\newcommand{\remark}{\underline{\sl Remark}: }

\newcommand{\HR}{{\cal H}}

\newcommand{\CB}{{\cal B}}

\newcommand{\FR}{{\cal F}}

\newcommand{\la}{\langle}
\newcommand{\ra}{\rangle}
\newcommand{\ga}{\gamma}

\newcommand{\dt}{\delta}
\newcommand{\Dt}{\Delta}
\newcommand{\vr}{\varepsilon}

\newcommand{\BR}{{\Bbb R}}

\newcommand{\Lb}{\Lambda}

\newcommand{\BJ}{{\Bbb J}}

\def\diag{\mathop{\rm diag}}

\def\be{\begin{equation}}
\def\ee{\end{equation}}
\def\bea{\begin{eqnarray}}
\def\eea{\end{eqnarray}}

\ifx\undefined\Bbb
        \let\Bbb\bf
\fi

\catcode`\@=11
\def\ps@X{\let\@mkboth\@gobbletwo
        \def\@oddhead{\tt Adler-van Moerbeke:%
        symmetric random\hfil\ \ August 24, 1998\ \ \hfil\S\thesection,
p.\thepage
        }
        \def\@oddfoot{\rm\hfil\thepage\hfil}
        \let\@evenhead\@oddhead
        \let\@evenfoot\@oddfoot}
\catcode`@=12
\catcode`\@=11
\let\c@equation=\relax
\newcounter{equation}[subsection]

\catcode`\@=12

\newtheorem{definition}{Definition}[
section]

\newtheorem{theorem}[definition]{Theorem}

\newtheorem{lemma}[definition]{Lemma}

\newtheorem{proposition}[definition]{Proposition}

\catcode`\@=11
\let\c@equation=\relax
\newcounter{equation}[
section]

\catcode`\@=12

\begin{document}
\maketitle


\setcounter{equation}{0}

\vspace{0.5cm}

\noindent Table of contents:
\medbreak
\noindent 0. Introduction
\newline\noindent 1. Borel decomposition and the 2-Toda lattice
\newline\noindent 2. Two-Toda $\tau$-functions and Pfaffian
$\TT$-functions
\newline \noindent 3. The Pfaffian Toda lattice and
skew-orthogonal polynomials
\newline\noindent 4. The ($s=-t$)-reduction of the
Virasoro vector fields
\newline\noindent 5. A representation of the Pfaffian $\tilde \tau$-function
as a symmetric matrix integral
\newline\noindent 6. String equations and Virasoro constraints
\newline\noindent 7. Virasoro constraints with boundary terms
\newline\noindent 8. Inductive Painlev\'e equations for Gaussian
and Laguerre ensembles
\newline\noindent 9. Appendix
\vspace{1cm}

The statistics governing the spectrum of Hermitean random matrix
ensembles is intimately related to the standard Toda lattice
\cite{AvM1}. The joint probabilities for the spectrum of coupled
Hermitean matrices are intimately related to the two-Toda lattice
\cite{AvM2,AvM3}. When the size of the matrices tend to $\iy$, the
probabilities involved in the bulk and edge scaling limits relate
to the Korteweg-de Vries equation; see \cite{ASV3,ASV4}. In his
doctoral dissertation, H. Peng \cite{P} shows, based on Mehta's
\cite{M1} pioneering work on the subject, that the symmetric matrix
models and the statistics of the spectrum of symmetric matrix
ensembles is governed by a peculiar reduction of the 2-Toda
lattice.

In each of these instances, the connection with the integrable
system is made by inserting in the probabilities above a
``time"-dependent exponential; in all our previous work, we
observed that the expressions thus obtained form a ratio of
$\tau$-functions for the associated integrable system.

What is the integrable system related to integrals over {\em
symmetric matrices}~? The purpose of this paper is to show that
Peng's reduction of the 2-Toda system does not lead to vectors of
$\tau$-functions, as he conjectured, but rather to a new vector of
function, which we shall call ``{\em Pfaffian $\tau$-functions}";
the vector actually lives in a {\em deeper stratum} for the
Birkhoff decomposition; neither do the individual Pfaffian
$\TT$-function obey the KP hierarchy, nor do they satisfy the
standard Toda bilinear equations . Instead, it is shown that $\TT$
satisfies a hierarchy of partial differential equations, different
from the KP-hierarchy (section 2 and \cite{ASV5}), and that, as a
whole, the vector of $\TT$'s gives rise to a Lax pair on matrices
(section 3), which we call the ``{\em Pfaffian Toda Lattice}" We
show the system satisfies Virasoro constraints (section 6 and 7),
reminiscent of the classical ones, as Peng's work suggests, leading
to inductive equations for the spectral probabilities (section 8) .

 Consider the hierarchy of equations
\begin{equation}
{\pl m_{\iy}\over\pl t_n}=\Lb^n m_{\iy},\quad\quad\quad {\pl
m_{\iy}\over\pl s_n}=-m_{\iy}\Lb^{\top n},\quad n=1,2,...,
\end{equation}
on bi-infinite matrices $m_{\iy}$ for skew-symmetric initial
condition $m_{\iy}(0,0)$, where the matrix
$\Lb=(\dt_{i,j-1})_{i,j\in\BZ}$ is the shift matrix. Then Borel
decomposing
$
m_{\iy}(t,s)
=S_1^{-1}S_2,~~\mbox{
for } ~t,s
\in
\BC^{\iy},
$
into lower- and upper-triangular matrices $S_1(t,s)$ and
$S_2(t,s)$, leads to a two-Toda system for $L_1:=S_1\Lb S^{-1}$ and
$L_2=S_2\Lb^{\top}S_2^{-1}$, which maintains the form
$m_{\iy}(t,s)=-m_{\iy}(-s,-t)^{\top}$. Its $\tau$-functions are
given by\footnote{$m_n:=(\mu_{ij})_{-\iy \leq i,j
\leq n-1}$} $\tau_n(t,s)=\det m_n(t,s)$. If the initial matrix
$m_{\iy}$ is semi-infinite (i.e., $\tau_0=1$), then
\be
\tau_n(t,s)=(-1)^n\tau_n(-s,-t).
\ee

When $s\rightarrow -t$, formula (0.3) shows that in the limit the
odd $\tau$-functions vanish, whereas the even $\tau$-functions are
determinants of skew-symmetric matrices. Thus, we are led to
considering {\em Pfaffians}:
\be
\TT_n(t):= \tau_n(t,-t)^{1/2}= (\det
~m_n(t,-t))^{1/2}, ~~\mbox{for even}~n\geq 0.
\ee
which satisfy a hierarchy of partial differential equations
\cite{ASV5}, reminiscent of the KP-hierarchy, but with the
 right hand side not equal to zero\footnote{$e^{\sum_1^{\iy}t_i
z^i}=\sum_0^{\iy}p_i(t)z^i,~~\tilde \pl=\left(\frac{\pl}{\pl
t_1},\frac{1}{2}\frac{\pl}{\pl t_2},\frac{1}{3}\frac{\pl}{\pl
t_3},...\right)$}:
\be
\left(p_{k+4}(\tilde\pl)-\frac{1}{2}\frac{\pl^2}{\pl
t_1\pl t_{k+3}}\right)\TT_{2n}\circ\TT_{2n}=p_k(\tilde
\pl)\TT_{2n+2}\circ\TT_{2n-2}
\ee
\noindent\hfill $k,n=0,1,2,...~.$ For $k=0$, equations (0.4)
can be viewed as an expression for $\TT_{2n+2}$ in terms of prior
$\TT$'s.

The {\em ``Pfaffian $\TT$-functions"}, themselves {\em not}
$\tau$-functions, tie up remarkably with the 2-Toda
$\tau$-functions, as follows\footnote{$[\al]:=(\al,\frac{\al^2}{2},
\frac{\al^3}{3},...)$}:
\bea
\tau_{2n}(t,-t-[\al])&=&\TT_{2n}(t)\TT_{2n}(t+[\al])\nonumber\\
& & \\
\tau_{2n+1}(t,-t-[\al])&=&-\al\TT_{2n}(t)
\TT_{2n+2}(t+[\al]).\nonumber
\eea
When $\al \rightarrow 0$, we are approaching a {\em deep} stratum
in the Borel decomposition of $m_{\iy}$, where every odd
$\tau$-function vanishes and hence the usual decomposition fails
.
When we approach the stratum according to to $[\al] \rightarrow 0$,
then one finds the formulae (0.5) above, established in
\cite{ASV5}. In particular, the odd $\tau$-functions
$\tau_{2n+1}(t,-t-[\al])$ approach zero linearly in $\al$.
Equations (0.5) are crucial in deriving bilinear relations from the
the 2-Toda lattice equations for Pfaffian $\TT$-functions
(mentioned in Theorem 2.2). They give rise to ``{\em
skew-orthogonal polynomials}", to ``{\em Pfaffian wave functions}"
and ultimately to Lax-type evolution equations on matrices $L$,
given by the AKS-theorem and to be explained in section 3:
$$
\frac{\pl L}{\pl
t_n}=\left[-\frac{1}{2}(L^n)_d-(L^n)_-+J(L^n)_+^{\top}J,L\right].
$$
Picking the matrix
\be
m_n(t,s):=\left(\mu_{ij}(t,s)\right)_{0\leq i,j\leq n-1}
\ee
of moments of the special type\footnote{$ \vr(x):=~\mbox{sign}~(x).
$}, for a subset $E\subset
\BR$,
\be
\mu_{k,\ell}(t,s)=\int\!\int_{E^{2}}
x^k y^{\ell} e^{V(x)+V(y)+\sum^{\iy}_
{i=1}(t_ix^i_k-s_iy^i)}\vr(x-y)dx dy,
\ee
yields an example of $m_{\iy}$ satisfying the hierarchy of
equations (0.1). Then we have
\bea
\tau_{\ell}(t,s)&=&\det m_{\ell}(t,s)\nonumber\\
& =&\int\!\int_{E^{2\ell}}
\prod^{\ell}_{k=1}\left(e^{V(x_k)+V(y_k)+\sum^{\iy}_
{i=1}(t_ix^i_k-s_iy_k^i)}\vr(x_k-y_k)\right)\Delta_{\ell}(x)
\Delta_{\ell}(y)\vec{dx}\vec{dy},\nonumber\\
\eea
and, with Peng \cite{P},
\be
\TT_{2n}(t)=\sqrt{\tau_{2n}(t,-t)}
=\int_{{\cal S}_{2n}(E)}e^{Tr(V(X)+\sum_1^{\iy}t_iX^i)}
dX,
\ee
for the Haar measure $dX$ on symmetric matrices and
$$
{\cal S}_{2n}(E):=\{2n\times 2n\mbox{\,\,symmetric matrices $X$
with spectrum $\in E\}$}.
$$
Assuming a potential $V$ and a disjoint union $E$ of the form,
\be
V'(z):=\frac{g}{f}=
\frac{\sum_0^{\iy} b_i z^i}{\sum_0^{\iy} a_i
z^i},\quad\quad
 E=\bigcup^r_{i=1}[c_{2i-1},c_{2i}]\subset \BR,
\ee
the integrals (0.9) satisfy the following Virasoro constraints:
\be
\left(\sum_{i=1}^{2r}c_i^{k+1}f(c_i)\frac{\pl}{\pl c_i}-\sum_{\ell=0}^{\iy}
\left(\frac{a_{\ell}}{2}\BJ_{k+\ell,N}^{(2)}+b_{\ell}
\BJ_{k+\ell+1,N}^{(1)}\right)\right)
\tilde\tau_{N}(t)=0
\ee
for all $k\geq -1$, and even $N\geq 0$,
where\footnote{in terms of
the customary Virasoro generators in $t_1,t_2,...$:
\begin{eqnarray*}
J_n^{(0)}&=&\dt_{n0},\hspace{2cm}J_n^{(1)}=\frac{\pl}{\pl
t_n}+(-n)t_{-n},\hspace{2cm} J_0^{(1)}=0\\
J_n^{(2)}&=&\sum_{i+j=n}:J_i^{(1)}J_j^{(1)}
:=\sum_{i+j=n}\frac{\pl^2}{\pl
t_i\pl t_j}+2\sum_{-i+j=n}it_i\frac{\pl}{\pl t_j}+
\sum_{-i-j=n}(it_i)(jt_j),
\end{eqnarray*}
}
\bea
\BJ^{(1)}_k&=&\left(\BJ^{(1)}_{k,n}\right)_{n\geq 0}=
\left(J_k^{(1)}+nJ_k^{(0)}\right)_{n\geq 0},\nonumber\\
\BJ_k^{(2)}&=&\left(\BJ^{(2)}_{k,n}\right)_{n\geq
0}=\left(J_k^{(2)}+(2n+k+1)J_k^{(1)}+n(n+1)J_k^{(0)}\right)_{n\geq
0}.\nonumber\\
\eea
As we have seen happen in all such cases, the ``{\em boundary}" and
``{\em time}" parts decouple! In particular, when $e^{V(z)}
\rightarrow 0$ sufficiently fast at the boundary of the set $E$,
the boundary differential operator is absent. This is also the
case, when $E=\BR$ and the integral (0.9) makes sense.

{\em What information does the integrable theory provide about the
probabilities
\be
P_{2n}(t,E):=\frac{\int_{{\cal
S}_{2n}(E)}e^{Tr(V(X)+\sum_1^{\iy}t_iX^i)} dX}{\int_{{\cal
S}_{2n}(\BR)}e^{Tr(V(X)+\sum_1^{\iy}t_iX^i)} dX},
\ee
after setting $t=0$~?} Upon expressing $\TT$'s partial derivatives
in $t$ at $t=0$ as partial derivatives in the $c_i$, by means of
the Virasoro constraints (0.11), and setting them in the non-linear
equations (0.4) for $\TT$, we find partial differential equations
in the $c_i$ for the $\TT$ and thus for the probabilities (0.13),
at $t=0$. Expressing partial derivatives in $t$ at $t=0$ in terms
of partial derivatives in the $c_i$ will only be possible, when the
potentials satisfy sufficiently strong conditions.

 For
example, for even $N$, the probability
\be
P_{N+2}(0,E)=\frac{\int_{{\cal S}_{N+2}(E)}e^{-Tr X^2}}
{\int_{{\cal S}_{N+2}(\BR)}e^{-Tr X^2}}~~~~~({\bf \mbox{Gaussian
integral}})
\ee
is expressed in terms of $P_{N}(0,E)$ and $P_{N-2}(0,E)$ and the
non-commutative operators
\be
\DR_k=\sum^{2r}_{i=1}c_i^{k+1}\frac{\pl}{\pl c_i},
\ee
 as follows:

\vspace{0.7cm}

$192~b_N\displaystyle
{\frac{P_{N+2}P_{N-2}}{P_N^2}}-12N(N-1)=$\hfill
\be
\hspace{1cm}(\DR_{-1}^4+8(2N-1)\DR_{-1}^2+12\DR_0^2+24\DR_0
-16\DR_{-1}\DR_1)\log P_N
+6(\DR^2_{-1}
\log P_N)^2.
\ee
When the set $E=[c, \iy]$, then  $P_{N+2}(0,E)$ is expressed in
terms of $G:=\frac{\pl}{\pl c}
\log
P_N(0,c)$, as follows:
\be
192 b_{N}\frac{P_{N+2}P_{N-2}}{P_N^2}-12N(N-1)
=G'''+6~G'^2-4\left(\,c^{2}-2\,(2N-1)\right)\,G'+4c~G,
\ee
 where the differential operator appearing on the right hand side
of (0.17) is reminiscent of the Painlev\'e IV equation.

In section 8, we also work out the PDE's for the probabilities for
the Laguerre ensemble:

\be
P_{N+2}(0,E)=\frac{\int_{{\cal S}_{N+2}(E)}e^{-Tr (X-\al \log X)}
dX} {\int_{{\cal S}_{N+2}(\BR^+)}e^{-Tr (X-\al \log X)}
dX}.~~~~~({\bf
\mbox{Laguerre integral}})
\ee

The paper contains two methods for obtaining the Virasoro
constraints (0.11). The most conceptual one is to establish string
relations. Indeed the Borel decomposition $m_{\iy}=S_1^{-1} S_2$
leads to so-called (monic) {\em string-orthogonal polynomials}
\cite{AvM3}:
\be
p^{(1)}(z)=:S_1\chi(z)~~\mbox{and}~~p^{(2)}(z)=:h
(S_2^{-1})^{\top}\chi(z),
\ee
satisfying orthogonality relations with respect to an inner
product\footnote{setting $h_n=\frac{\det m_{n+1}}{\det
m_{n}}=\frac{\tau_{n+1}}{\tau_n}$}$$
\la p_n^{(1)}(z),p_m^{(2)}(z)\ra
 =\dt_{m,n}h_n~,
$$
determined by the weight appearing in formula (0.8). They also play
an important role in developping the skew-othogonal polynomials,
alluded to earlier.

Acting with $z$ and $\pl/\pl z$ on the semi-infinite vectors
$p^{(i)}(z)$ leads to semi-infinite matrices $L_i$ and $M_i$
respectively, for $i=1,2$. Assuming the potential $V$ of the form
(0.10) and a sufficiently fast decay to $0$ of $e^{V(z)}$ at the
boundary of its support, we have that the orthogonality of the
polynomials leads to ``{\em string equations}" for $k\geq
-1$:
\bigbreak
\noindent $ M_1L_1^{k+1}f(L_1)-M_2L_2^{k+1}f(L_2)$\hfill
\be
+L_1^{k+1}g(L_1)+L_2^{k+1}g(L_2)+(L_1^{k+1}
f(L_1))'+L_2^kf(L_2)
 =0.
\ee
The ASV-correspondence \cite{ASV1} enables one to translate the
string equations (0.20) into {\em Virasoro constraints} for the
2-Toda $\tau$-functions $\tau_{\ell}$ and, in turn, for the
Pfaffian $\TT$-functions.

\section{Borel decomposition and the 2-Toda lattice}

\setcounter{equation}{0}

In \cite{AvM4,AvM2}, we considered the following differential
equations for the bi-infinite or semi-infinite moment matrix
$m_{\iy}$
\begin{equation}
{\pl m_{\iy}\over\pl t_n}=\Lb^n m_{\iy},\quad {\pl m_{\iy}\over\pl
s_n}=-m_{\iy}\Lb^{\top n},\quad n=1,2,...,
\end{equation}
where the matrix $\Lb=(\dt_{i,j-1})_{i,j\in\BZ}$ is the shift
matrix; then (1.1) has the following solutions:
\begin{equation}
m_{\iy}(t,s)=e^{\sum t_n\Lb^n}m_{\iy}(0,0)e^{-\sum s_n\Lb^{\top
n}}\end{equation} in terms of some initial condition
$m_{\iy}(0,0)$.

Consider  the Borel decomposition $m_{\iy}=S_1^{-1}S_2$, for
\begin{eqnarray*}
S_1\in G_-&=&\{\mbox{lower-triangular invertible matrices, with
1's on the diagonal}\}\\
S_2\in G_+&=&\{\mbox{upper-triangular invertible matrices}\},
\end{eqnarray*}
with corresponding Lie algebras $g_-,g_+$; then setting
$L_1:=S_1\Lb S^{-1}$,
\begin{eqnarray*}
S_1\frac{\pl m_{\iy}}{\pl
t_n}S_2^{-1}&=&S_1(S_1^{-1}S_2)^.S_2^{-1}=-\dot S_1S_1+\dot
S_2S_2^{-1}\in g_-+g_+\\ &=&S_1\Lb^n m_{\iy}
S_2^{-1}=S_1\Lb^nS_1^{-1}=L_1^n=(L_1^n)_-+(L_1^n)_+\in g_-+g_+;
\end{eqnarray*}
the uniqueness of the decomposition $g_-+g_+$ leads to
$$
-\frac{\pl S_1}{\pl t_n}S_1^{-1}=(L_1^n)_-,\quad\frac{\pl S_2}{\pl
t_n}S_2^{-1}=(L_1^n)_+.
$$
Similarly setting $L_2=S_2\Lb^{\top}S_2^{-1}$, we find
$$
-\frac{\pl S_1}{\pl s_n}S_1^{-1}=-(L_2^n)_-,\quad
\frac{\pl S_2}{\pl s_n}S_2^{-1}=-(L_2^n)_+.
$$
This leads to the 2-Toda equations for $S_1,S_2$ and $L_1,L_2$:
\be
\frac{\pl S_{1,2}}{\pl t_n}=\mp(L_1^n)_\mp S_{1,2},\quad
\frac{\pl S_{1,2}}{\pl s_n}=\pm(L_2^n)_\mp S_{1,2}
\ee

\be
\frac{\pl L_i}{\pl t_n}=[(L_1^n)_+,L_i],\quad
\frac{\pl L_i}{\pl s_n}=[(L_2^n)_-,L_i],\quad i=1,2,...
\ee
By 2-Toda theory \cite{AvM4} the problem is solved in terms of a
sequence of tau-function
$$\tau_n(t,s)=\det m_n(t,s),
$$
with $m_n(t,s)$ defined in (0.6). Notice that in the

\noindent \underline{bi-infinite case} ($n \in \BZ$):
$$m_n(t,s):=\left(\mu_{ij} (t,s)\right)_{-\iy<i,j\leq
n-1},$$

\noindent \underline{semi-infinite case} ($n\geq 0$):
\be
m_n(t,s):=\left(\mu_{ij}(t,s)\right)_{0\leq i,j\leq
n-1},~~\mbox{with}~\tau_0=1.
\ee

The two pairs of wave functions $\Psi=(\Psi_1,\Psi_2)$ and
$\Psi^{\ast}=(\Psi_1^{\ast},\Psi_2^{\ast})$ defined by
$$
\Psi_{1}(t,s,z)=e^{\sum_1^{\iy} t_i z^{
i}}S_{1}\chi(z),~~\Psi^{\ast}_{1}(t,s,z)
=e^{-\sum_1^{\iy} t_i z^{
i}}\left(S^{\top}_{1}\right)^{-1}\chi(z^{-1})
$$
\be
\Psi_{2}(t,s,z)=e^{\sum_1^{\iy} s_i z^{-
i}}S_{2}\chi(z),~~\Psi^{\ast}_{2}(t,s,z)
= e^{-\sum_1^{\iy} s_i z^{-
i}}\left(S^{\top}_{2}\right)^{-1}\chi(z^{-1})
\ee
 satisfy
 $$
L\Psi=(z,z^{-1})\Psi,\quad
 L^*\Psi^*=(z,z^{-1})\Psi^*,
 $$
 and
$$
\left\{
\begin{array}{l}
\frac{\pl }{\pl t_n}\Psi=((L_1^n)_+,(L_1^n)_+)\Psi\\
\frac{\pl }{\pl s_n}\Psi=((L_2^n)_-,(L_2^n)_-)\Psi
 \end{array}
\right.
$$

\begin{equation}
\left\{
\begin{array}{l}
\frac{\pl}{\pl t_n}\Psi^* =-((L_1^n)_+,(L_1^n)_+)^{\top}
\Psi^{\ast}\\
\frac{\pl}{\pl s_n}\Psi^*
=-(((L_2^n)_-,(L_2^n)_-)^{\top}\Psi^*.
\end{array}
\right.
\end{equation}
Also define the wave operators
\be
W_1:=S_1e^{\sum t_i \Lb^i} ~~\mbox{and}~~W_2:=S_2 e^{\sum s_i
\Lb^{\top i}},
\ee
and the operators $M_i$ and $M_i^{\ast}$
\footnote{$\vr$ and $\vr^{\ast}$ are infinite matrices such that
$\vr\chi(z)=\frac{\pl}{\pl z}\chi(z), $ and
$\vr^*\chi(z)=\frac{\pl}{\pl z^{-1}}\chi(z). $ }:
$$
M=(M_1,M_2):=(W_1\vr W_1^{-1},W_2\vr^* W_2^{-1})
$$
\be
M^*=(M_1^*,M_2^*)=(-M_1^{\top}+L_1^{\top
-1},-M_2^{\top}+L_2^{\top -1}).
\ee
The operators $L,L^*,M$ and $M^*$ satisfy, in view of (1.1):
\be
\begin{array}{c}
        L\Psi=(z,z^{-1})\Psi,\quad
        M\Psi=\Bigl({\pl\over\pl z},{\pl\over\pl(z^{-1})}\Bigr)\Psi,\quad
        [L,M]=(1,1),
        \\
 L^*\Psi^*=(z,z^{-1})\Psi^*,\quad
        M^*\Psi=\Bigl({\pl\over\pl z},
            {\pl\over\pl(z^{-1})}\Bigr)\Psi^*,\quad
        [L^*,M^*]=(1,1).
        \\[5mm]

\end{array}
\ee

In \cite{UT}, with a slight notational modification \cite{ASV2} the
wave functions are shown to have the following $\tau$-function
representation:
\bea
\Psi_1(t,s;z)&=&\biggl(
        \frac{\tau_n(t-[z^{-1}],s)}{\tau_n(t,s)}
e^{\sum^{\iy}_1 t_iz^i}z^n
\biggr)_{n\in\BZ} \nonumber\\
\Psi_2(t,s;z)&=&\biggl(
        \frac{\tau_{n+1}(t,s-[z])}{\tau_n(t,s)}
e^{\sum^{\iy}_1 s_iz^{-i}}z^n
\biggr)_{n\in\BZ} \nonumber\\
\Psi^*_1(t,s,z)&=&\Biggl(\frac{\tau_{n+1}(t+[z^{-1}],s)}
{\tau_{n+1}(t,s)}e^{-\sum^{\iy}_1
t_iz^i}z^{-n}\Biggr)_{n\in\BZ} \nonumber\\
\Psi^*_2(t,s,z)&=&\Biggl(\frac{\tau_n(t,s+[z])}
{\tau_{n+1}(t,s)}e^{-\sum^{\iy}_1
s_i z^{-i}}z^{-n}\Biggr)_{n\in\BZ},
\eea
with the following bilinear identities satisfied for the wave
and adjoint wave functions $\Psi$ and $\Psi^{\ast}$,
for all
$m,n\in\BZ$ (bi-infinite) and
$m,n\geq 0$ (semi-infinite) and $t,s,t',s'\in\BC^{\iy}$:
\begin{equation}
\oint_{z=\iy} \Psi_{1n}(t,s,z) \Psi_{1m}^{\ast} (t',s,',z)
\frac{dz}{2 \pi
i z}=\oint_{z=0} \Psi_{2n}(t,s,z) \Psi_{2m}^{\ast}(t',s,',z)
\frac{dz}{2
\pi i z}.
\end{equation}
The $\tau$-functions\footnote{The first contour runs
clockwise about a small neighborhood of $z=\iy$, while the
second runs counter-clockwise about $z=0$.} satisfy the
following bilinear identities:
$$
\oint_{z=\iy}\tau_n(t-[z^{-1}],s)\tau_{m+1}(t'+[z^{-1}],s')
e^{\sum_1^{\iy}(t_i-t'_i)z^i}
z^{n-m-1}dz
$$
\be
=\oint_{z=0}\tau_{n+1}(t,s-[z])\tau_m(t',s'+[z])
e^{\sum_1^{\iy}(s_i-s'_i)z^{-i}}z^{n-m-1}dz;
\ee
they characterize the 2-Toda lattice $\tau$-functions. Note (1.6)
and (1.11) yield
\be
(S_2)_0=\diag(...,\frac{\tau_{n+1}(t,s)}{\tau_{n}(t,s)},...):=
h(t,s).
\ee

The symmetry vector fields $\BY_N$ acting on $\Psi$ and $L$,
$$
\BY_{M_i^{\al}L_i^{\beta}}\left(\Psi_1,\Psi_2 \right):=(-1)^{i-1}
\left(-(M_i^{\al}L_i^{\beta})_{-}\Psi_1,
(M_i^{\al}L_i^{\beta})_+\Psi_2\right),
$$
$$
\BY_{M_i^{\al}L_i^{\beta}}\left(L_1,L_2\right):=
(-1)^{i-1}\left(\left[-(M_i^{\al}L_i^{\beta})_{-},L_1\right],
\left[(M_i^{\al}L_i^{\beta})_+,L_2\right]\right)
$$
for $i=1,2$ and $\al,\beta \in \BZ, \al\geq 0$, lift to an action
on $\tau$, according to the Adler-Shiota-van Moerbeke formula
\cite{ASV1,ASV2}:

\begin{proposition}  For $n,k\in\BZ$, $n\geq 0$,
the symmetry vector fields $\BY_{M_i^nL_i^{n+k}}$,\linebreak $(i=1,2)$ acting on
$\Psi$ lead to the correspondences\footnote{where
$$
        \eta=\sum_1^\iy{z^{-i}\over i}{\pl\over\pl t_i}
        \quad\hbox{and}\quad
        \tilde\eta=\sum_1^\iy{z^i\over i}{\pl\over\pl s_i},
$$
so that
$$
        e^{a\eta+b\tilde\eta}f(t,s)=f(t+a[z^{-1}],s+b[z])
$$}
\begin{eqnarray}
-{((M_1^nL_1^{n+k})_{-}\Psi_1)_m\over\Psi_{1,m}}
&=& {1\over n+1}
(e^{-\eta}-1){W_{m,k}^{(n+1)}(\tau_m)\over\tau_m},\nonumber\\
{((M_1^nL_1^{n+k})_{+}\Psi_2)_m\over\Psi_{2,m}} &=& {1\over
n+1}\biggl(
e^{-\tilde\eta}{W_{m+1,k}^{(n+1)}(\tau_{m+1})\over\tau_{m+1}}
-{W_{m,k}^{(n+1)}(\tau_m)\over\tau_m}
\biggr),\nonumber\\
{((M_2^nL_2^{n+k})_{-}\Psi_1)_m\over\Psi_{1,m}} &=& {1\over n+1}
(e^{-\eta}-1){\tilde W_{m-1,k}^{(n+1)}(\tau_m)
\over\tau_m},\nonumber\\
-{((M_2^nL_2^{n+k})_{+}\Psi_2)_m\over\Psi_{2,m}}
&=&
{1\over n+1}\biggl(
e^{-\tilde\eta}{\tilde W_{m,k}^{(n+1)}(\tau_{m+1})\over\tau_{m+1}}
-{\tilde W_{m-1,k}^{(n+1)}(\tau_m)\over\tau_m}
\biggr),\nonumber\\
\end{eqnarray}
\end{proposition}

In Proposition 1.1, the $W$-generators take on the following form
in terms of the customary $W$-generators
\be
W_{n,\ell}^{(k)}=\sum_{j=0}^k\MAT{1}n\\j\mat(k)_jW_{\ell}^{(k-j)}
\quad\mbox{and}\quad
\tilde W_{n,\ell}^{(k)}=W_{-n,\ell}^{(k)}\Bigl|_{t\rg s}.
\ee
We shall only need the $W_{n,\ell}^{(k)}$- generators for $0\leq k
\leq 2$:
\be
W_n^{(0)}=\dt_{n,0},\quad W_n^{(1)}=J_n^{(1)}\quad\mbox{and}\quad
W_n^{(2)}=J_n^{(2)}-(n+1)J_n^{(1)},\quad\quad n\in\BZ
\ee
and
\be
\begin{array}{lllllll}
W_{m,i}^{(1)}&=W_i^{(1)}+mW_i^{(0)}&&&&W_{m,i}^{(2)}&=W_i^{(2)}+2mW_i^{(1)}
+m(m-1)W_i^{(0)}\\
 &=J_i^{(1)}+m\dt_{i0}&&&& &=J_i^{(2)}+(2m-i-1)J_i^{(1)}
+m(m-1)\dt_{i0},
\end{array}
\ee
expressed in terms of the Virasoro generators $J$ (see footnote 5).
The corresponding expression $\tilde W^{(k)}_{m,i}$ can be read off
from the above, using (1.16), with $J_n^{(k)}$ replaced by $\tilde
J_n^{(k)}
= J_{n}^{(k)}\mid_{t\rg s}$.

\section{Two-Toda $\tau$-functions and Pfaffian
$\TT$-functions}

In this section we state the properties of the 2-Toda lattice,
associated with an initial skew-symmetric bi-infinite matrix
$m_{\iy}(0,0)$, which the reader can find in \cite{ASV5}. When the
matrix $m_{\iy}(0,0)$ is semi-infinite, the $\tau$-functions
$\tau_n(t,s)$ have the property
\be
\tau_n(t,s)=(-1)^n\tau_n(-s,-t).
\ee

\begin{theorem}{\em \cite{ASV5}} If the initial matrix
$m_{\iy}(0,0)$ is skew-symmetric, then, under the 2-Toda flow,
$m_{\iy}(t,s)$ evolves as follows:
\be
m_{\iy}(t,s)=-m_{\iy}(-s,-t)^{\top}.
\ee
Moreover,
$$
h^{-1}S_1(t,s)=-S_2^{\top
-1}(-s,-t)\quad\mbox{and}\quad h^{-1}S_2(t,s)=S_1^{\top -1}(-s,-t),
$$
$$
h^{-1}\Psi_1(t,s,z)=-\Psi_2^*(-s,-t,z^{-1})\quad\mbox{and}\quad
h^{-1}\Psi_2(t,s,z)=\Psi_1^*(-s,-t,z^{-1}),
$$
$$
L_1(t,s)=h L_2^{\top}h^{-1}(-s,-t)\quad\mbox{and}\quad L_2(t,s)=h
L_1^{\top}h^{-1}(-s,-t),
$$
\be
\mbox{with}\quad h(-s,-t)=- h(t,s).
\ee
Finally in the semi-infinite case
\be
\tau_n(-s,-t)=(-1)^n\tau_n(t,s).
\ee
\end{theorem}

For a skew-symmetric semi-infinite initial matrix $m_{\iy}(0,0)$,
relation (2.2) guarantees the skew-symmetry of $m_{\iy}(t,-t)$.
Therefore the odd $\tau$-functions vanish and the even ones have a
natural square root, the {\em Pfaffian} $\TT_{2n}(t)$:
\be
\tau_{2n+1}(t,-t)=0,\quad\tau_{2n}(t,-t)=:\tilde\tau_{2n}
^2(t),\ee
where the Pfaffian, together with its sign specification, is
determined by the formula:
\bea
\TT_{2n}(t)dx_0\wedge dx_1\wedge ...\wedge dx_{2n-1}&:=&
pf \left(m_{2n}\right)dx_0\wedge dx_1\wedge ...\wedge
dx_{2n-1}\nonumber\\ &=&\frac{1}{n!}\left(\sum_{0\leq i<j\leq
2n-1}\mu_{ij}(t,-t)dx_i\wedge dx_j\right)^n.\nonumber\\
\eea

\begin{theorem}{\em \cite{ASV5}}  For a semi-infinite, skew-symmetric initial condition
$m_{\iy}(0,0)$, the 2-Toda $\tau$-function $\tau(t,s)$ and the
Pfaffians $\TT(t)$ are related by
\bea
\tau_{2n}(t,-t-[\al])&=&\TT_{2n}(t)\TT_{2n}(t+[\al])\nonumber\\
& & \nonumber\\
\tau_{2n+1}(t,-t-[\al])&=&-\al\TT_{2n}(t)\TT_{2n+2}(t+[\al])
\nonumber
\eea
or alternatively
\bea
\tau_{2n}(t-[\al],-t)&=&\TT_{2n}(t-[\al])\TT_{2n}(t)\nonumber\\
& & \\
\tau_{2n+1}(t-[\al],-t)&=&-\al\TT_{2n}(t-[\al])\TT_{2n+2}(t).\nonumber
\eea
The $\TT$-functions satisfy the bilinear relations
\bea
&
&\oint_{z=\iy}\TT_{2n}(t-[z^{-1}])\TT_{2m+2}(t'+[z^{-1}])
e^{\sum_0^{\iy}(t_i-t'_i)z^i}
z^{2n-2m-2}dz\nonumber\\
&+&\oint_{z=0}\TT_{2n+2}(t+[z])\TT_{2m}(t'-[z])
e^{\sum_0^{\iy}(t'_i-t_i)z^{-i}}z^{2n-2m}dz=0.
\eea
\end{theorem}

 The next theorem states that the Pfaffian
$\TT$-functions satisfy differential Fay identities, but also
hierarchy of equations, whose left hand side is reminiscent of the
KP-equation, but augmented with a non-zero right hand side.

\begin{theorem}{\em \cite{ASV5}}
For semi-infinite, skew-symmetric initial condition $m_{\iy}(0,0)$,
the functions $\TT_{2n}(t)$ satisfy the following ``differential
Fay identity"\footnote{$\{f,g\}=f'g-fg'$, where $'=\pl/\pl t_1$.}

\bigbreak

$
\{\TT_{2n}(t-[u]),\TT_{2n}(t-[v])\}
$\hfill
\bea
& & \hspace{2cm}+(u^{-1}-v^{-1})(\TT_{2n}(t-[u])\TT_{2n}(t-[v])
-\TT_{2n}(t)\TT_{2n}(t-[u]-[v]))\nonumber\\
& & \nonumber\\
 & &~~~=uv(u-v)\TT_{2n-2}(t-[u]-[v])\TT_{2n+2}(t),
\eea
and Hirota type bilinear equations, always involving nearest
neighbours:
\be
\left(p_{k+4}(\tilde\pl)-\frac{1}{2}\frac{\pl^2}{\pl
t_1\pl t_{k+3}}\right)\TT_{2n}\circ\TT_{2n}=p_k(\tilde
\pl)~\TT_{2n+2}\circ\TT_{2n-2}
\ee
\hfill $k,n=0,1,2,...~.$
\end{theorem}

\section{The Pfaffian Toda lattice and skew-\\orthogonal polynomials}

Consider the $(t,s)$-dependent inner product
$$
\la f,g \ra=\int\!\int_{\BR^2}f(x) g( y)
e^{\sum^{\iy}_1(t_ix^i-s_iy^i)} F(x,y)dx\,dy,~~t,s \in \BC^{\iy},
$$
with regard to the skew-symmetric weight $F(x,y)$,
\be
F(y,x)=-F(x,y),
\ee
and the moment matrix
\be
m_n(t,s):=\left(\mu_{k\ell}(t,s)\right)_{0\leq k,\ell\leq
n-1}=\left(\la x^k , y^{\ell}\ra \right)_{0\leq k,\ell\leq n-1}.
\ee
Define
\bea
\tau_n(t,s)
&:=&\det  m_n (t,s) \nonumber\\ & & \nonumber
\\ &=&\int\!\int_{(\BR^2)^n}\prod^n_{k=1}
\left(e^{\sum^{\iy}_{i=1}(t_i
x_k^i-s_iy^i_k)} F(x_k,y_k)\right)
\Dt_n(x)\Dt_n(y)\vec{dx}\,\vec{dy}.\nonumber\\
\eea
The proof that the two expressions (3.3) for $\tau_n$ are
identical, is based on an identity involving Vandermonde
determinants and can be found in \cite{AvM2}.

We are exactly in the framework of section 2; indeed, on the one
hand, the moments $\mu_{ij}$ in (3.2) satisfy the equations
\be
\frac{\pl\mu_{ij}}{\pl t_k}=\mu_{i+k,j}\quad\mbox{and}\quad
\frac{\pl\mu_{ij}}{\pl s_k}=-\mu_{i,j+k},
\ee
and so $m:=m_{\iy}$ satisfies (0.1), and $(\tau_n(t,s))_{n\geq 0}$
is a 2-Toda $\tau$-vector.

On the other hand, the skew-symmetry (3.1) of $F$ implies
$\mu_{ij}(0,0)=-\mu_{ji}(0,0)$, and so the skewness of
$m_{\iy}(0,0)$; so, by theorem 2.1, we have
$$
\mu_{ij}(s,t)=-\mu_{ji}(-s,-t).
$$
Therefore also, $m_{\iy},S_1,S_2,\Psi_1,\Psi_2,h$ and $\tau$ have
the properties mentioned in theorem 2.1, and
\be
\TT_{2n}(t)=\left( \det m_{2n}(t,-t) \right)^{1/2}=pf(0,...,2n-1)
\ee
satisfies the relations of Theorem 2.2 and the non-linear hierarchy
, mentioned in Theorem 2.3.

We now construct the vector of ``{\em wave functions}"
$\tilde\Psi=(\tilde\Psi_k)_{k\geq 0}$, containing the functions
$p:=(p_k)_{k\geq 0}$
:
\begin{eqnarray}
\tilde\Psi_{2n}(t,z)&:=&e^{\sum t_i z^i}\tilde p_{2n}(t,z)
:=e^{\sum t_i z^i}z^{2n}\frac{\TT_{2n}(t-[z^{-1}])}
{\TT_{2n}(t)}\nonumber\\
\tilde\Psi_{2n+1}(t,z)&:=&e^{\sum t_i z^i}
\tilde p_{2n+1}(t,z):=e^{\sum t_i z^i}\frac{\left(z+\frac{\pl}{\pl
t_1}\right)z^{2n}\TT_{2n}(t-[z^{-1}])}{\TT_{2n}(t)},\nonumber\\
\end{eqnarray}
and the diagonal matrix
\be
\tilde
h=\mbox{diag}~(\tilde h_0,\tilde h_0,\tilde h_1,\tilde
h_1,...)\mbox{ with }
\tilde h_n:=\frac{\TT_{2n+2}(t)}{\TT_{2n}(t)}.\ee

Consider the lie algebra of semi-infinite matrices, whose elements
are $2\times 2$ matrices, instead of scalars. The $0$th band
consists of $2\times 2$ matrices along the diagonal, whereas the
first band contains $2\times 2$ matrices just above the $0$th band,
etc... This leads to a gradation of matrices $\DR=\sum_i \DR_i,
~\DR_+=\sum_{i > 0}
\DR_i,~~\DR_-=\sum_{i < 0}\DR_i$, giving rise to the matrix decomposition
$A=A_-+A_0+A_+$. For future use, define the matrix $J=\sqrt{-I} \in
\DR_0$,
\be J:=
\left(
\begin{array}{cc|cc|cc}
 0&1& & & &\\
-1&0& & & &\\
\hline
  & &0&1& &\\
 & &-1&0& &\\
\hline
 & & & & &\\
\end{array}\right),~~~~\mbox{with}~ J^2=-I.
\ee

\begin{theorem}
The $\tilde p_k(t,z)$ are monic polynomials\footnote{with the
understanding that $pf($odd set$)=0$ and $pf(0,...,\widehat
k,...,\widehat{2n},2n+1)=-pf(0,...,2n-1)$ for $k=2n+1$.} in $z$:
\bea
\tilde p_{2n}(t,z)&=&\frac{1}{\TT(t)} \sum_{0\leq k\leq
2n}(-z)^k pf(0,...,\hat k,...,2n,\widehat{2n+1})\nonumber\\ &=&
z^{2n}+\sum_{0\leq k\leq 2n-1}(-z)^k
\frac{pf(0,...,\hat k,...,2n)}{pf(0,...,2n-1)}
\nonumber\\
&&\nonumber\\
 \tilde p_{2n+1}(t,z)&=&\frac{1}{\TT(t)}
 \sum_{0\leq k\leq 2n+1}(-z)^k pf(0,...,\hat
k,...,\widehat{2n},2n+1) \nonumber\\
 &=& z^{2n+1}+
 \sum_{0\leq k\leq 2n-1}(-z)^k \frac{pf(0,...,\hat
k,...,\widehat{2n},2n+1)}{pf(0,...,2n-1)} ,
\eea
which are skew-orthogonal with respect to the inner product
$\la~,~\ra|_{s=-t}$,
$$
\left\{\begin{array}{l}
\la\tilde p_{2n}(z),\tilde p_{2n+1}(z)\ra|_{s=-t} =-\la\tilde
p_{2n+1}(z),\tilde p_{2n}(z)\ra |_{s=-t}
=\frac{\TT_{2n+2}(t)}{\TT_{2n}(t)}=\tilde h_n\\
\\
\la\tilde p_i(z),\tilde p_j(z)\ra|_{s=-t} =0\quad \mbox{otherwise,}
\end{array}
\right. ;
$$
i.e.,
\be
(\la\tilde p_i,\tilde p_j\ra)_{0\leq i,j<\iy}=J\tilde h=:\tilde J
.\ee

\end{theorem}

The polynomials $\tilde p_{2n}(z)$ give rise to the semi-infinite
matrix $\tilde P$ of coefficients of the polynomials $\tilde
p_{2n}(z)$; i.e.,
\be
\tilde P\chi(z)=\tilde p(z),
\ee
which, in view of (3.9), has the form
$
\tilde P\in I+\DR_-,
$ taking into account the precise definition of $\DR_-$. We now
dress up the shift $\Lb$ by means of $\tilde P$, yielding two
matrices:
\be
\tilde L:=\tilde P\Lb\tilde P^{-1}\quad\quad\mbox{and}\quad\quad
L:=\tilde h^{-1/2}\tilde L\tilde h^{1/2}.\ee Define the projection
\begin{eqnarray*}
\HR:\DR&\longrightarrow&\DR_0\\
A&\longmapsto& A_d:=~A+JA^{\top}J\\ & & \hspace{0.61cm}=\mbox{
diag}(a_{00}+a_{11},a_{00}+a_{11}, a_{22}+a_{33},a_{22}+a_{33},...)
.
\end{eqnarray*}
With this notation, we now state:

\begin{theorem} ({\bf Pfaffian Toda Lattice}) The matrices $\tilde L$ and $L$ satisfy the
equations for $n\geq 1$,
\be
\frac{\pl \tilde L}{\pl t_n}=[-(\tilde L^n)_-+\tilde
J(\tilde L^n)_+ ^{\top}\tilde J~,\tilde L]
\ee
and
\be\frac{\pl L}{\pl
t_n}=\left[-\frac{1}{2}(L^n)_d-(L^n)_-+J(L^n)_+^{\top}J,L\right].
\ee
\end{theorem}

\section{The ($s=-t$)-reduction of the Virasoro vector fields}

 In this section we explain, as a remarkable feature, how the Virasoro
vector fields for 2-Toda behave under the reduction $s=-t$.

\begin{proposition} $\tau_{2n}(t,s)$ satisfies the following identity,
near $s=-t$, for $i=1,2$:
\be
\left(J^{(i)}_{\ell}+(-1)^i\tilde J^{(i)}_{\ell}\right)\tau_{2n}(t,s) |_{s=-t}
=2 \sqrt{\tau_{2n}(t,-t)}\hat J^{(i)}_{\ell}
\left( \sqrt{\tau_{2n}(t,-t)}\right),
\ee
where $\hat J^{(i)}_{\ell}$ on the right hand side is the same
operator $J^{(i)}_{\ell}$, but with partials $\pl/\pl t_i$ replaced
by total derivatives $d/dt_i$.
\end{proposition}

\noindent The proof is based on identities, involving skew-symmetric
matrices and Pfaffians. Indeed to a skew-symmetric matrix
$A_{2n-1}$ augmented with an arbitrary row and column
$$
M=\left(
\begin{array}{ccc|c}
 & & &x_0\\
 &A_{2n-1}& & \\
 & & & \\
 & & &x_{2n-2}\\
\hline
-y_0&...&-y_{2n-2}&z
\end{array}\right),
$$
we associate, in a natural way, skew-symmetric matrices
$$
A=\left(
\begin{array}{ccc|c}
 & & &x_0\\
 &A_{2n-1}& &\vdots\\
 & & & \\
 & & &x_{2n-2}\\
\hline
 &\ast& &0\end{array}\right)\quad\quad B=\left(
\begin{array}{ccc|c}
 & & &y_0\\
 &A_{2n-1}& &\vdots\\
 & & & \\
 & & &y_{2n-2}\\
\hline
 &\ast& &0
\end{array}\right).
$$

Similarly, to a skew-symmetric matrix $A_{2n-2}$ augmented
with two arbitrary rows and columns
$$
N=\left(
\begin{array}{cccc|cc}
 & & & &x_0&y_0\\
 & &A_{2n-2}& &\vdots&\vdots\\
 & & & &x_{2n-3}&y_{2n-3}\\
\hline
-u_0&...& &-u_{2n-3}&-u_{2n-2}&y_{2n-2}\\
-v_0&...& &-v_{2n-3}&x_{2n-1}&-v_{2n-1}
\end{array}\right),
$$
we associate the four skew-symmetric matrices
$$
C=\left(
\begin{array}{ccc|cc}
 & & &x_0&v_0\\
 &A_{2n-2}& &\vdots&\vdots\\
 & & &x_{2n-3}&v_{2n-3}\\
\hline
 & & &0&-x_{2n-1}\\
 &\ast& & & \\
 & & &x_{2n-1}&0
\end{array}\right)\quad\quad D=\left(
\begin{array}{ccc|cc}
 & & &u_0&y_0\\
 &A_{2n-2}& &\vdots&\vdots\\
 & & &u_{2n-3}&y_{2n-3}\\
\hline
 & & &0&y_{2n-2}\\
 &\ast& & & \\
 & & &-y_{2n-2}&0
\end{array}\right)
$$
$$
E=\left(
\begin{array}{ccc|cc}
 & & &x_0&u_0\\
 &A_{2n-2}& &\vdots&\vdots\\
 & & &x_{2n-3}&u_{2n-3}\\
\hline
 & & &0&u_{2n-2}\\
 &\ast& & & \\
 & & &-u_{2n-2}&0
\end{array}\right)\quad\quad F=\left(
\begin{array}{ccc|cc}
 & & &v_0&y_0\\
 &A_{2n-2}& &\vdots&\vdots\\
 & & &v_{2n-3}&y_{2n-3}\\
\hline
 & & &0&-v_{2n-1}\\
 &\ast& & & \\
 & & &v_{2n-1}&0
\end{array}\right)
$$


\begin{lemma} Given the matrices $M$ and $N$ above, we have
\begin{eqnarray*}
\det M&=&pf(A)pf(B)\\
\det N&=&pf(C)pf(D)-pf(E)pf(F).
\end{eqnarray*}
\end{lemma}

\bigbreak

In the proof of proposition 4.1, we shall use the following
symbolic notation for the Pfaffian of $\mu$:
$$
pf(0,...,2n-1):= \sqrt{\det \left(\mu_{ij}\right)}_{0\leq i,j \leq
2n-1},
$$
with the sign specification mentioned in (2.6).

\bigbreak

\noindent\underline{\sl Proof of Proposition 4.1}: Since
$$
\tau_{2n}(t,s)=\tau_{2n}(-s,-t),
$$
we have
\be
\left(\frac{\pl}{\pl t_i}+\frac{\pl}{\pl s_i}\right)
\tau_{2n}\Bigl|_{s=-t}
=0,~~\left(\frac{\pl^2}{\pl s_i\pl t_j}-
\frac{\pl^2}{\pl s_j\pl t_i}\right)
\tau_{2n}\Bigl|_{s=-t}
=0.
\ee
Since $J_{\ell}^{(2)}$ for $\ell \geq -1$ consists of two parts, a
first order one and a second order one for $\ell\geq 2$, let us
first dispose of the first one:

\medbreak

$\displaystyle{\sum^{\iy}_{i=1}\left(it_i\frac{\pl}{\pl
t_{i+\ell}}+ is_i\frac{\pl}{\pl
s_{i+\ell}}\right)\tau_{2n}(t,s)\Biggl|_{s=-t}}$
\begin{eqnarray}
&=&\sum^{\iy}_{i=1}it_i\left(\frac{\pl}{\pl t_{i+\ell}}-
\frac{\pl}{\pl s_{i+\ell}}\right)\tau_{2n}(t,s)\Biggl|_{s=-t}\nonumber\\
&=&\sum^{\iy}_{i=1}it_i\frac{d}{d
t_{i+\ell}}\tau_{2n}(t,-t)\nonumber\\
&=&2\sqrt{\tau_{2n}(t,-t)}\left(\sum^{\iy}_{i=1}it_i
\frac{\pl}{\pl t_{i+\ell}}\right)\sqrt{\tau_{2n}(t,-t)}.
\end{eqnarray}

 To deal with the second order part of $J_{\ell}^{(2)}$,
 with $\ell \geq 2 $ and
 setting $\tau_{2n}(t,s)=f^2(t,s)$, we compute:
\medbreak

$\displaystyle{\sum_{i+j=k}\left\{\left(\frac{\pl^2}{\pl t_i\pl t_j}+
\frac{\pl^2}{\pl s_i\pl s_j}\right)\tau_{2n}(t,s) \Biggl|_{s=-t}
-2\tau_{2n}(t,-t)^{1/2}\frac{d^2}
{dt_idt_j}\tau_{2n}(t,-t)^{1/2}\right\}}$
\begin{eqnarray}
&=&\displaystyle{\sum_{i+j=k}\left\{\left(\frac{\pl^2}{\pl t_i\pl
t_j}+
\frac{\pl^2}{\pl s_i\pl s_j}\right)f^2(t,s)\Biggl|_{s=-t}-2f(t,-t)\frac{d^2}
{dt_idt_j}f(t,-t)\right\}}\nonumber\\
 &=&2\sum_{i+j=k}\Biggl\{\frac{\pl
f}{\pl t_i}\frac{\pl f}{\pl t_j}+
\frac{\pl f}{\pl s_i}\frac{\pl f}{\pl s_j}+f\frac{\pl^2f}
{\pl t_i\pl t_j}+f\frac{\pl^2f}{\pl s_i\pl s_j}\nonumber \\ &
&\hspace{2cm}-f\left(\frac{\pl}{\pl t_j}-\frac{\pl}{\pl s_j}\right)
\left(\frac{\pl}{\pl t_i}-
\frac{\pl}{\pl s_i}\right)f\Biggr\}_{s=-t}  \nonumber   \\
&=&2\sum_{i+j=k}\Biggl\{\frac{\pl f}{\pl t_i}\frac{\pl f}{\pl t_j}+
\frac{\pl f}{\pl s_i}\frac{\pl f}{\pl s_j}+f\frac{\pl^2 f}
{\pl t_j\pl s_i}+f\frac{\pl^2f}{\pl s_j\pl
t_i}\Biggr\}_{s=-t}\nonumber\\
&=&\frac{1}{2}\sum_{i+j=k}\Biggl\{4\left(\frac{\pl}{\pl t_i}
-\frac{\pl}{\pl s_i}\right)f
\left(\frac{\pl}{\pl t_j}-\frac{\pl}{\pl s_j}\right)f  \nonumber \\
& &\hspace{2cm}+4\frac{\pl}{\pl s_i}\left(f\frac{\pl f}{\pl t_j}
\right)+4
\frac{\pl}{\pl t_i}
\left(f\frac{\pl f}{\pl s_j}\right)\Biggr\}_{s=-t}\nonumber\\
&=&\frac{1}{2f^2}\sum_{i+j=k}\Biggl\{\left(\frac{\pl}{\pl t_i}
-\frac{\pl}{\pl s_i}\right)f
^2\left(\frac{\pl}{\pl t_j}-\frac{\pl}{\pl s_j}\right)f^2+2f^2\left(\frac{\pl^2}
{\pl s_i\pl t_j} +\frac{\pl^2}{\pl t_i\pl
s_j}\right)f^2\Biggr\}_{s=-t}\nonumber\\
&=&\frac{2}{f^2}\sum_{i+j=k}\Biggl\{\frac{\pl f^2}{\pl
t_i}\frac{\pl f^2}{\pl t_j} +f^2 \frac{\pl^2 f^2}{\pl t_i\pl
s_j}\Biggr\}_{s=-t}~~\mbox{using (4.2)}\nonumber\\
&=&2\sum_{i+j=k}\Biggl\{\frac{\pl \tau_{2n}/\pl t_i}{
\sqrt{\tau_{2n}}}\frac{\pl
\tau_{2n}/\pl t_j}{\sqrt{\tau_{2n}}} +\frac{\pl^2 \tau_{2n}}
{\pl t_i \pl s_j}\Biggr\}_{s=-t},~~\mbox{using}~
f^2=\tau_{2n}(t,s)\nonumber\\ &=& 0
\end{eqnarray}

The vanishing of this last expression, is based on the argument
below, using (3.4). Indeed the action of $\frac{\pl}{\pl t_i}$
(respectively, $\frac{\pl}{\pl s_j}$) on the determinant of the
moment matrix $m_{2n}$ amounts to a sum (over $0\leq k \leq 2n-1$)
of determinants of the same matrices, but with the $k^{{\rm th}}$
row (respectively, the $k^{{\rm th}}$ column) replaced by
$(\mu_{k+i,0},...,\mu_{k+i,2n-1})$ (respectively, by
$(\mu_{0,\ell+j},...,\mu_{2n-1,\ell+j})^{\top}$). Thus, the
matrices in the sum are 
matrices of size $2n-1$, which are skew-symmetric except for an
additional (arbitrary) row and column. Note that, when $0\leq
k+i\leq 2n$, the corresponding matrix has zero determinant. So,
using the first relation of Lemma 4.2, one
finds\footnote{$pf(0,...,\ell\mapsto
\ell+i,...,n-1)$ denotes the Pfaffian of the skew-symmetric
matrix $m_n(t,-t)$, with the $\ell$th column replaced by the
$\ell+i$th column of $m_{\iy}(t,-t)$.}

\bea
\frac{\pl \tau_{2n}}
{\pl t_i }\Biggl|_{s=-t} &=&\sum_{\ell} pf(0,...,2n-1)~
pf(0,...,\ell\mapsto
\ell+i,...,2n-1)\nonumber\\ &=&\sqrt{\tau_{2n}(t,-t)}~\sum_{\ell}pf(0,...,\ell\mapsto
\ell+i,...,2n-1),\nonumber
 \eea
and hence,

\vspace{0.5cm} $
\displaystyle{\frac{\frac{\pl \tau_{2n}}{\pl t_i}}{
\sqrt{\tau_{2n}}}\frac{\frac{\pl
\tau_{2n}}{\pl t_j}}{\sqrt{\tau_{2n}}}\Biggl|_{s=-t}}$\hfill
\be=\sum_{\ell,m} pf(0,...,m \mapsto m+j,...,2n-1)
pf(0,...,\ell\mapsto
\ell+i,...,2n-1) \}
\ee

Similarly, the second derivative $\pl^2/\pl s_i\pl t_j$ amounts to
a sum of determinants (over $0\leq m,\ell \leq 2n-1$) of
skew-symmetric matrices, except that the $\ell$th row and $m$th
column got replaced by the $\ell +i$th row and $m+j$th column
respectively. So, all in all, we get a sum of determinants
 of the second type
in Lemma 4.2, thus leading to

\vspace{0.5cm}
$
\displaystyle{-\frac{\pl^2 \tau_{2n}}
{\pl t_i \pl s_j}\Biggl|_{s=-t} }$\hfill
\bea
&=&\sum_{\ell,m}\det (\ell\mbox{th row}~ \mapsto (\ell +i)\mbox{th
row},~~m\mbox{th column}~ \mapsto (m+j) \mbox{th column})
\nonumber\\ &=&\sum_{\ell,m}
\{pf(...,m\mapsto m,...,\ell
\mapsto
\ell,...) pf(...,m\mapsto m+j,...,\ell \mapsto
\ell+i,...)\nonumber\\ & & +~pf(...,m\mapsto m,...,\ell \mapsto
m+j,...) pf(...,m\mapsto
\ell+i,...,\ell \mapsto
\ell,...) \}\nonumber\\
& & \nonumber\\
 &=&\sum_{\ell,m} \{pf(0,...,n-1) pf(0,...,m\mapsto
m+j,...,\ell \mapsto
\ell+i,...,2n-1)\nonumber\\ & & +~pf(0,...,\ell \mapsto
m+j,...,2n-1) pf(0,...,m\mapsto
\ell+i,...,2n-1) \}.\nonumber\\
&&
 \eea
Therefore, summing both contributions (4.5) and (4.6), one finds:
$$
-\sum_{i+j=k}\Biggl\{\frac{\pl \tau_{2n}/\pl t_i}{
\sqrt{\tau_{2n}}}\frac{\pl
\tau_{2n}/\pl t_j}{\sqrt{\tau_{2n}}} +\frac{\pl^2 \tau_{2n}}
{\pl t_i \pl s_j}\Biggr\}_{s=-t}\hspace{5cm}
$$
\bea
&=&\sum_{\ell,m,i+j=k}  pf(0,...,m\mapsto m+j,...,\ell
\mapsto
\ell+i,...,2n-1) ~pf(0,...,2n-1)\nonumber\\
&+&\sum_{\ell,m,i+j=k}\Bigl\{ pf(0,...,\ell \mapsto m+j,...,2n-1)
pf(0,...,m\mapsto
\ell+i,...,2n-1)\nonumber\\
&&~~~- ~pf(0,...,m \mapsto m+j,...,n-1)~ pf(0,...,\ell\mapsto
\ell+i,...,2n-1) \Bigr\}\nonumber\\
\eea

The expression above consists of two sums; we now show each of the
sums vanish separately. The first sum vanishes, because it is a sum
of zero pairs\footnote{A Pfaffian flips sign, upon permuting two
indices.}
$$
pf(...,m\mapsto m+j,...,\ell
\mapsto
\ell+i,...)+pf(...,m\mapsto m+j',...,\ell
\mapsto
\ell+i',...)=0,
$$
upon picking $m+j'=\ell+i,~\ell+i'=m+j$, thus respecting the
requirement $i+j=i'+j'=k$. The argument is similar for the second
sum in (4.7).\qed

\section{A representation of the Pfaffian $\tilde \tau$-function
as a symmetric matrix integral}

In this section we consider skew-symmetric weights (3.1) of the
special form:
\be
F(x,y):=e^{V(x)+V(y)}I_E(x)I_E(y) \vr(x-y),
\ee
for a union of intervals $E
\subset
\BR$.
Here we give a matrix representation, due to Peng \cite{P}, based
on arguments of Mehta \cite{M1}, for
$\TT_{2n}=\tau_{2n}(t,-t)^{1/2}$ in terms of the set
$$
{\cal S}_{2n}(E):=\{2n\times 2n\mbox{\,\,symmetric matrices $X$
with spectrum $\in E\}$}.
$$

\bigbreak

\begin{theorem} If
\bea
\tau_{\ell}(t,s)&=&\det
\left(\mu_{ij}(t,s)\right)_{0\leq ij\leq \ell-1} \nonumber\\
&=&\int\!\int_{E^{2\ell}}
\prod^{\ell}_{k=1}\left(e^{V(x_k)+V(y_k)+\sum^{\iy}_
{i=1}(t_ix^i_k-s_iy_k^i)}\vr(x_k-y_k)\right)\Delta_{\ell}(x)
\Delta_{\ell}(y)\vec{dx}\vec{dy},\nonumber\\
\eea
 then
$$
\TT_{2n}(t)=\sqrt{\tau_{2n}(t,-t)}
=\int_{{\cal S}_{2n}(E)}e^{Tr\left(V(X)+\sum_1^{\iy}t_iX^i\right)}
dX,
$$
where $dX$ is Haar measure on symmetric matrices.
\end{theorem}

\proof Upon setting $V(x,t):=V(x)+\sum_0^{\iy}t_i
x^i $, and
$$
F_i(x):=\int^x_{-\iy}y^ie^{V(y,t)}I_E(y)dy ,\quad\mbox{and}
\quad G_i(x):=x^ie^{V(x,t)}I_E(x),
$$
we notice the following representation for the moment:
\bea
\mu_{ij}(t,-t)&=&\int\!\int_{\BR^2}x^iy^je^{V(x,t)+
V(y,t)}\vr(x-y)I_E(x)I_E(y)dx\,dy\nonumber\\ &=&\int\!\int_{x\geq
y}(x^iy^j-x^jy^i)e^{V(x,t)+V(y,t)}I_E(x)I_E(y)dx\,dy\nonumber\\
&=&\int_E(F_j(x)G_i(x)-F_i(x)G_j(x))dx.
\eea

Applying the spectral theorem to the symmetric matrix $X=
\\O^{\top}$
 diag$(x_1,...,x_{2n})O$, with $O\in SO(2n)$, we find
\be
dX=|\Delta_{2n}(x)|dx_1...dx_{2n}\,dO.
\ee
Upon integrating the orthogonal group, one finds:
\bigbreak
\noindent$
\int_{{\cal S}_{2n}(E)}e^{Tr V(X,t) } dX
$\hfill
\begin{eqnarray*}
&=&c_{2n}\int_{\BR^{2n}}\prod_1^{2n}(e^{V(x_i,t)}
I_E(x_i)dx_i)|\Delta_{2n}(x)|\\
&=&c'_{2n}\int_{-\iy<x_1<x_2<...<x_{2n}<\iy}\det(x_j^i
e^{V(x_j,t)}I_E(x_j)dx_j)_{\begin{tabular}{l} $0\leq i\leq 2n-1$\\
$1\leq j\leq 2n$
\end{tabular}},\\
&&\hspace{8cm}\mbox{setting} ~c'_{2n}=(2n)! c_{2n}\\
&=&c'_{2n}\int_{-\iy<x_2<x_4<...<x_{2n}<\iy}
\prod^n_{k=1}(dx_{2k} e^{V(x_{2k},t)}I_E(x_{2k}))\\ & &
\det\left(F_i(x_2)~~x_2^i~~F_i(x_4)-F_i(x_2)~~
x_4^i~...~F_i(x_{2n})-F_i(x_{2n-2})~~x^i_{2n}\right)_{0\leq i\leq
2n-1}\\ & &\\
&=&c'_{2n}\int_{-\iy<x_2<x_4<...<x_{2n}<\iy}
\prod^n_1(dx_{2k} e^{V(x_{2k},t)}I_E(x_{2k}))
\\
& &\hspace{2cm} \det\left(F_i(x_2)~~~x_2^i~~~F_i
(x_4)~~~x_4^i~~...~~F_i(x_{2n})~~~x_{2n}^i\right)_{0\leq i\leq
2n-1}\\ & &
\\
&=&c'_{2n}\int_{-\iy<x_1<x_2<...<x_n<\iy}\prod^n_1dx_i\\ & &
\hspace{3cm}\det\Bigl(F_i(x_1)~~G_i(x_1)~~...~~F_i (x_n)~~G_i(x_n)\Bigr)_{0\leq
i\leq 2n-1}\\ &=&\frac{c_{2n}}{n!}\int_{\BR^n}\prod^n_1dx_i~~
\det\Bigl(F_i(x_1)~~G_i(x_1)~~...~~F_i(x_n)~~
G_i(x_n)\Bigr)_{0\leq i\leq 2n-1},\\ & &\hspace{8cm}\mbox{upon
permuting the $x_i$},\\
&=&c'_{2n}\Bigl(\det\left(G_i(x)F_j(x)-F_i(x) G_j(x)\right)_{0\leq
i,j\leq n-1}\Bigr)^{1/2},\hspace{1cm}\\ & &\mbox{\hspace{6cm} using
de Bruijn's Lemma \cite{M1},p.446},\\
&=&c'_{2n}\det\left(\mu^E_{ij}(t,-t)\right)^{1/2}_{0\leq i,j\leq
n-1}
\mbox{\,\,using (5.2)}\\
&=&c_{2n}\TT_{2n}(t),
\end{eqnarray*}
using (2.5).
\qed

\section{String equations and Virasoro constraints}

In this section, we consider the moments (3.2), with regard to the
skew-symmetric weight
\be
F(x,y):=e^{V(x)+V(y)} \vr(x-y),
\ee
assuming the following form for the potential $V$, as in (0.10):
\be
V'(z)=\frac{g}{f}=\frac{\sum_0^{\iy}b_i z^i}{\sum_0^{\iy}a_i z^i},
\ee
with $e^{V(z)}$ decaying to $0$ fast enough at the boundary of its
support.

 According to \cite{AvM2,AvM4}, the
semi-infinite matrices $S_1=S_1(t,s)\in\DR_{-\iy,0}$ and $ S_2=
S_2(t,s)\in\DR_{0,\iy}$ in the Borel decomposition $
m_{\iy}=S_1^{-1}S_2 $ of the semi-infinite moment matrix $m_{\iy}$
lead to string-orthogonal (monic) polynomials
\be
p^{(1)}(z)=:S_1\chi(z)~~\mbox{and}~~p^{(2)}(z)=:h
(S_2^{-1})^{\top}\chi(z),
\ee
satisfying the orthogonality relations
$$
\la p_n^{(1)},p_m^{(2)}\ra =\dt_{n,m}h_n
$$
for the inner product
\be
\la f,g \ra = \int_{\BR^2 } dy dz~ \vr(y-z)
e^{V(y)+V(z)+\sum_1^{\iy}(t_i y^i-s_i z^i)} f(y) g(z).
\ee

\vspace{0.5cm}
Besides $L_1\in\DR_{-\iy,1}$ and $L_2\in\DR_{-1,\iy}$, we also
define $Q_1, Q_2\in\DR_{-\iy,-1}$, as follows
\be
\begin{array}{lll}
{\rm (i)}&zp_n^{(1)}(z)=\sum_{\ell\leq
n+1}(L_1)_{n\ell}p_{\ell}^{(1)}(z)& zp_n^{(2)}(z)=\sum_{\ell\leq
n+1}(h L_2^{\top}h^{-1})_{n\ell}p_{\ell}^{(2)}(z)\\
 & & \\
{\rm (ii)}&\frac{\pl}{\pl z}p_n^{(1)}(z)=\sum_{\ell\leq
n-1}(Q_1)_{n\ell}p_{\ell}^{(1)}(z)&\frac{\pl}{\pl
z}p_n^{(2)}(z)=\sum_{\ell\leq n-1}(
Q_2)_{n\ell}p_{\ell}^{(2)}(z).\\
\end{array}
\ee
Finally, setting for the sake of this section,
$$
V_t(x):=\sum_1^{\iy}t_i x^i~~\mbox{and}~~V_s(z):=\sum_1^{\iy}s_i
y^i,
$$
we define matrices
\be
M_1:=Q_1+\frac{\pl V_{t}}{\pl y}(L_1),~~~M_2:=- h  Q^{\top}_2
h^{-1}+\frac{\pl V_s}{\pl z}(L_2)+L_2^{-1},
\ee
which are shown to be compatible with the definition of the $M_i$
in (1.9). We now state the two main theorems of this section:

\begin{theorem} {\rm (``String equations")}.
The semi-infinite matrices $L_i$ and $M_i$ satisfy the following
matrix identities in terms of $~V'=g/f$, for all $k\geq
-1$:

\bigbreak

\noindent $M_1L_1^{k+1}f(L_1)-M_2L_2^{k+1}f(L_2)$\hfill
\be
+L_1^{k+1}g(L_1)+L_2^{k+1}g(L_2)+(L_1^{k+1}
f(L_1))'+L_2^kf(L_2)
 =0.
\ee
\end{theorem}

\vspace{0.7cm}

The proof of this theorem will be postponed until later in this
section. This fact, together with the ASV-correspondence
(Proposition 1.1) and proposition 5.1, leads at once to the
constraints for the 2-Toda $\tau$-functions and the Pfaffian
$\TT$-functions.

\begin{theorem} {\rm (``Virasoro constraints")}. The multiple integrals
($\tau_0=1$)
\bea
\tau_n(t,s)&=&\det\left(\mu_{ij}(t,s)\right)_{0\leq i,j\leq n-1}
\nonumber \\
&=&\int\!\int_{\BR^{2n}}\prod^n_{k=1}\left(e^{V(x_k)+V(y_k)+\sum^{\iy}_
{i=1}(t_ix^i_k-s_iy_k^i)}\vr(x_k-y_k)\right)\Delta_n(x)
\Delta_n(y)\vec{dx}\vec{dy}  \nonumber
\eea
form a $\tau$-vector for the 2-Toda lattice and satisfy the
following Virasoro constraints for all $k\geq -1$ and $n\geq 0$:
\be
\sum_{i\geq 0}
\left\{\begin{array}{l}
\frac{a_i}{2}\left(J^{(2)}_{i+k}+\tilde
J^{(2)}_{i+k}+(2n+i+k+1)(J^{(1)}_{i+k}-\tilde
J_{i+k}^{(1)})+2n(n+1)J^{(0)}_{i+k}\right)\\
 \\
+ b_i\left(J^{(1)}_{i+k+1}-\tilde
J^{(1)}_{i+k+1}+2nJ^{(0)}_{i+k+1}\right)
\end{array}
\right\}\tau_n=0.
\ee
The Pfaffian
$$
\TT_N(t)=\tau_N(t,-t)^{1/2}=\int_{\BR^N}\prod_{k=1}^N
\left(e^{V(x_k)+\sum_i^{\iy}t_i x_k^i}\right)|\Dt_N(x)|\vec{dx},
~~N~\mbox{even},
$$
 satisfies the
Pfaffian-KP hierarchy (2.9) and (2.10), together with the following
Virasoro constraints, for all $k\geq -1$ and even $N\geq 0$:
\be
\sum_{\ell=0}^{\iy}
\left(\frac{a_{\ell}}{2}\BJ_{k+\ell,N}^{(2)}+b_{\ell}
\BJ_{k+\ell+1,N}^{(1)}\right)
\tilde\tau_{N}(t)=0.
\ee
\end{theorem}

 Before proceeding with the proof of theorems 6.1
and 6.2, we must show that the matrices $L_i$ and $M_i$, obtained
by (6.5) and (6.6), coincide with the ones defined in the general
theory, as in (1.9).

\begin{lemma} 
The string-orthogonal polynomials relate to the wave vectors
$\Psi_1$ and $\Psi^*_2$ as follows:
\be \begin{tabular}{lll}
$\Psi_1:=e^{\sum t_kz^k}p^{(1)}(z)$&\mbox{and}& $\Psi_2^*:=e^{-\sum
s_k z^{-k}}h^{-1}p^{(2)}(z^{-1}) $   \\ $\,\,\,~~~=e^{\sum t_k
z^k}S_1\chi(z)$& &$\,\,\,~~~=e^{-\sum
s_kz^{-k}}(S_2^{-1})^{\top}\chi(z^{-1}),$
\end{tabular}
\ee
whereas the matrices $L_1,~L_2^*=L_2^{\top}, ~M_1,
M_2^*=(L_2^{-1}-M_2)^{\top}$ satisfy the desired relations
$$(L_1,L_2^{\ast})(\Psi_1,\Psi_2^*)=(z,z^{-1})
(\Psi_1,\Psi_2^*)
$$
\be
(M_1,M_2^*)(\Psi_1,\Psi_2^*):=\Bigl(\frac{\pl}{\pl
z},\frac{\pl}{\pl z^{-1}}\Bigr)(\Psi_1,\Psi_2^*) .
\ee
\end{lemma}

\medbreak

\proof  Indeed, $\Psi_1$ and $\Psi_2^*$,
defined in (6.10), have the correct asymptotics. Also, we check,
using (6.5), that
$$
z\Psi_1=e^{\sum t_kz^k}zp^{(1)}(z)=e^{\sum
t_kz^k}L_1p^{(1)}=L_1\Psi_1,
$$
and
\bea
z^{-1}\Psi_2^*(z)=e^{-\sum s_kz^{-k}}h^{-1}z^{-1}
p^{(2)}(z^{-1})&=& e^{-\sum s_kz^{-k}}
L^{\top}_2 h^{-1}
p^{(2)}(z^{-1}))\nonumber\\&=& L_2^{\top}\Psi_2^*,\nonumber
\eea
leading to the first formula (6.11). From the second formula for
$\Psi_1$ and $\Psi_2^*$, one shows that
$$L_1=S_1\Lb S_1^{-1}~~\mbox{and}~~
L_2=  S_2\Lb^{\top}  S_2^{-1}.$$

To prove the second formula (6.11), observe, from (6.5), that for
$\Psi_1=\Psi_1(t,s;z)$
\begin{eqnarray*}
M_1\Psi_1=\frac{\pl\Psi_1}{\pl z}
&=&\frac{\pl}{\pl z}\left(e^{\Sg t_kz^k}p^{(1)}(z)\right)\\
&=&\sum_{k\geq 1}kt_kz^{k-1}\Psi_1+e^{\Sg t_iz^k}\frac{\pl}{\pl z}p^{(1)}(z)\\
&=&\Bigl(\frac{\pl V_t}{\pl z}(L_1)+Q_1\Bigr)\Psi_1,
\end{eqnarray*}
and similarly that for $\Psi^*_2=\Psi_2^*(t,s ;z)$,
\begin{eqnarray*}
M_2^{\ast}\Psi_2^{\ast}=\frac{\pl\Psi_2^*}{\pl z^{-1}}
&=&\frac{\pl}{\pl z^{-1}}\left(e^{-\sum s_k
z^{-k}}h^{-1}p^{(2)}(z^{-1})\right)\\ &=&-\sum k s_k
z^{-k+1}\Psi_2^*+e^{-\sum s_kz^{-k}}h^{-1}\frac{\pl}{\pl
z^{-1}}p^{(2)}(z^{-1})\\
&=&\Bigl(-\frac{\pl V_s(L_2^{\top})}{\pl z}+h^{-1}
Q_2h\Bigr)\Psi_2^*, \end{eqnarray*} concluding the proof of Lemma
6.3.\qed

\medbreak

\underline{\sl Proof of Theorem 6.1}: Using
$$
\frac{\pl}{\pl y}\vr(y-z)=2\delta(y-z),$$
setting $V_t(x)=\sum^{\iy}_1t_ix^i$, and using the hypothesis that
$e^V$ vanishes fast enough at the boundary of its
support\footnote{We imagine doing the calculation for all $t_i$ and
$s_j$ vanishing beyond $t_{2k}$ and $s_{2k}$ and letting the latter
be strictly negative and positive respectively.}, we compute, at
first,
\begin{eqnarray*}
0&=&\int_{\BR}dy\frac{\pl}{\pl
y}y^k f(y)\left\{\left(\int_{\BR}dz~\vr(y-z)e^{V(z)-V_s(z)}p_m^{(2)}
(z)\right)e^{V(y) +
V_t(y)}p_n^{(1)}(y)\right\}\\
&=&\int_{\BR}dy  \left(\int_{\BR}dz~\vr(y-z) e^{V(z)-V_s(z)}p_m^{(2)}
(z)\right)~e^{V(y) +
V_t(y)}\\
& &\hspace{1cm}\left\{\left((V(y)+V_t(y))'f(y)y^k
+(y^kf(y))'\right)p_n^{(1)} (y)+p_n^{(1)'}(y)y^k
f(y)\right\}\\
& &   \\
&&\hspace{0.5cm}+2\int
\int_{\BR^2}e^{V(y)
+
V(z)+V_t(y)-V_s(z)}y^k f(y)
 p_n^{(1)}
(y) p_m^{(2)}(z)\dt(y-z) dy dz \\
& &  \\
&=&\int\int_{\BR^2}dy dz \vr(y-z)e^{V(y)+V(z)+V_t(y)-V_s(z)}
\\
& & \hspace{0.7cm} \left\{\left(g(L_1)
L_1^k+(L_1^kf(L_1))'+(V'_t(L_1)+Q_1)L_1^kf(L_1)\right)p^{(1)}(y)
\right\}_n p_m^{(2)}(z) \\
& &  \hspace{0.5cm}
+2\int_{\BR}e^{2 V(y)+V_t(y)-V_s(y)}p_n^{(1)}(y)
p_m^{(2)}(y)y^k f(y)dy \\
& &  \\
&=&\left\{(Q_1+V'_t(L_1))L_1^k
f(L_1)+g(L_1)L_1^k+(L_1^kf(L_1))'
\right\}_{nm}h_m \\
&&\hspace{0.5cm}+2\int_{\BR}e^{2 V(y)+V_t(y)-V_s(y)}p_n^{(1)}(y)
p_m^{(2)}(y)y^k f(y)dy.
\end{eqnarray*}
Next, setting $\bar L_2:=h L_2^{\top} h^{-1}$, so that
$zp^{(2)}=(\bar L_2 p^{(2)})_n$, we find
\begin{eqnarray*}
0&=&\int_{\BR}dz\frac{\pl}{\pl z}z^k f(z)\left\{\left(\int_{\BR} dy
\vr(y-z) e^{V(y)+V_t(y)} p_n^{(1)}(y)\right) e^{V(z)-V_s(z)}
p_m^{(2)}(z)\right\}\\ & &   \\ &=&\int_{\BR}dz \left(\int_{\BR}
p_n^{(1)}(y)e^{V(y)+V_t(y)}\vr(y-z)dy\right)e^{V(z)-V_s(z)}\\ & &
\hspace{0.5cm}
\left\{\left((V(z)-V_s(z))'f(z)z^k+(z^k f(z))'\right)
p_m^{(2)}(z)+ p_m^{(2)'}(z)z^k f(z)
\right)\\
& &    \\
& &\hspace{0.5cm}-2\int
\int_{\BR^2}e^{V(y)
+ V(z)+V_t(y)-V_s(z)}f(z)z^k
 p_n^{(1)}
(y) p_m^{(2)}(z)\dt(y-z) dy dz\\ & &   \\
&=&  \int\int_{\BR^2}dy
dz ~ \vr(y-z)  e^{V(y) + V(z)+V_t(y)-V_s(z)}  \\ & & \hspace{1cm}
\left\{\left( ( g(\bar L_2) \bar L_2^k+ (f(\bar L_2) \bar
L_2^k)'+( Q_2-V_s'(\bar L_2))f(\bar L_2)\bar
L_2^k\right)p^{(2)}(z)\right\}_mp_n^{(1)}(y)\\ &
&-2\int_{\BR^2}e^{2V(y)+V_t(y)-V_s(y)}f(y)y^kp_n^{(1)}(y)
p_m^{(2)}(y) dy \\ & &   \\ &=&\left\{( Q_2-V'_s(\bar L_2))\bar
L_2^k f(\bar L_2)+g(\bar L_2)\bar L_2^k+(\bar L_2^k f(\bar L_2))'
\right\}_{mn}h_n\\
& & \hspace{0.5cm} -2
\int_{\BR^2}e^{2 V(y)
+
V_t(y)-V_s(y)}f(y)y^k
 p_n^{(1)}
(y) p_m^{(2)}(y) dy.
\end{eqnarray*}

\vspace{0.3cm}
\noindent Adding the two expressions yields the matrix identity

\vspace{0.2cm}
$\{(Q_1+V'_t(L_1))L_1^kf(L_1)+g(L_1)L_1^k+(L_1^kf(L_1))'\}h$\hfill
\be
+h\{( Q_2-V'_s(\bar L_2))\bar L_2^kf(\bar L_2)+g(\bar L_2)\bar
L_2^k+ (\bar L_2^kf(\bar L_2))'\}^{\top}=0.
\ee
Upon shifting $k\rg k+1$, upon using
$$
h^{-1}\bar L_2h=L_2^{\top},~~~Q_1+V'_t(L_1)=M_1, ~~~(h^{-1}
Q_2h)^{\top}-V'_s(L_2)=L_2^{-1}-M_2,$$ we have that identity (6.12)
leads to
\begin{eqnarray*}
M_1L_1^{k+1}f(L_1)&+&g(L_1)L_1^{k+1}+\left(L_1^{k+1}f(L_1)\right)'\\
&+&L_2^{k+1}f(L_2)(L^{-1}_2-M_2)+L_2^{k+1}g(L_2)
+\left(L_2^{k+1}f(L_2)\right)'=0;
\end{eqnarray*}
Finally, the fact that for any function $F(z)$, (since
$[L_2,M_2]=I$)
$$ F(L_2)M_2=M_2F(L_2)+F'(L_2),$$ leads to the identity,
 announced in theorem 6.1.\qed

\noindent\underline{{\em Proof of Theorem 6.2}}:
 Using the representation (6.2) of $V'(z)$, one obtains
 from (6.7) that
\begin{eqnarray*}
& &\sum_{i\geq 0}a_i(M_1L_1^{k+i+1}-M_2L_2^{k+i+1}+
(i+k+1)L_1^{i+k}+L_2^{i+k})\\ &&\hspace{3cm}+\sum_{i\geq
0}b_i(L_1^{i+k+1}+L_2^{i+k+1})=0.
\end{eqnarray*}
We now apply proposition 1.1.  The vanishing of the matrix
expression above implies obviously that the $(~)_-$ and $(~)_+$
parts vanish as well, so that acting respectively on the wave
vectors $\Psi_1$ and $\Psi_2$ lead to the vanishing of the four
right hand sides of (1.15) in proposition 1.1, for the
corresponding combination of $W$'s. Therefore we have

\medbreak

\noindent $\LR_{k,m}\tau_m$
\begin{eqnarray*}
&:=&\sum_{i\geq 0}\left\{\begin{array}{l}
a_i(W^{(2)}_{m,k+i}+\tilde
W^{(2)}_{m-1,k+i}+2(i+k+1)W^{(1)}_{m,i+k}-2\tilde
W^{(1)}_{m-1,i+k})\\
 \\
+2 b_i(W^{(1)}_{m,k+i+1}-W^{(1)}_{m-1,k+i+1})
\end{array}
\right\}\tau_m  \\
&&\\
 &=&c_k\tau_m;
\end{eqnarray*}
the point is that $c_k$ is independent of $t$, using the first and
third relations of proposition 1.1, and independent of $s$ and $n$
using the second and fourth relations. Finally, in view of the
relations (1.18), we have
\medbreak

\noindent $\LR_{k,m}\tau_m$
\begin{eqnarray*}
&=&\Biggl\{\sum_{i\geq 0}a_i\Bigl(J^{(2)}_{i+k}+\tilde
J^{(2)}_{i+k}+(2m-i-k-1)J^{(1)}_{i+k}\\ &
&\hspace{3cm}+(2(1-m)-i-k-1)\tilde
J_{i+k}^{(1)}+2m(m-1)\delta_{i+k,0}\Bigr)\\ & &\quad +2\sum_{i\geq
0}a_i\left((i+k+1)(J^{(1)}_{i+k}+m\delta_{i+k,0})-(\tilde
J^{(1)}_{i+k}+(1-m)\delta_{i+k,0})\right)\\ & &\quad +2\sum_{i\geq
0}b_i\left((J^{(1)}_{i+k+1}+m\delta_{i+k+1,0})-(\tilde
J^{(1)}_{i+k+1}+(1-m)\delta_{i+k+1,0})\right)\Biggr\}\tau_m\\
&=&\Biggl\{\sum_{i\geq 0}a_i\left(J^{(2)}_{i+k}+\tilde
J^{(2)}_{i+k}+(2m+i+k+1)(J^{(1)}_{i+k}-\tilde
J^{(1)}_{i+k})+(2m(m+1)-2)\delta_{i+k,0}\right)\\ & &\quad +2\sum
b_i\left((J^{(1)}_{i+k+1}-\tilde
J^{(1)}_{i+k+1})+(2m-1)\delta_{i+k+1,0}\right)\Biggr\}\tau_m.
\end{eqnarray*}
Since $c_k$ is independent of $m$ and $\tau_0=1$, and since most of
$\LR_{k,m}$ vanishes, when acting on a constant, we have
$$
\frac{\LR_{k,m}\tau_m}{\tau_m}=\frac{\LR_{k,0}\tau_0}{\tau_0}=-2\sum_{i\geq
0}(a_i\delta_{i+k,0}+b_i\delta_{i+k+1,0}),
$$
and so $$\left(\LR_{k,m}+2\sum_{i\geq
0}(a_i\delta_{i+k,0}+b_i\delta_{i+k+1,0})\right)\tau_m=0,$$
yielding the identity (6.8). The proof of the Virasoro constraints
$$
\sum_{i\geq 0}
\left\{\begin{array}{l}
\frac{a_i}{2}\left(J^{(2)}_{i+k}+(2n+i+k+1)J^{(1)}_{i+k}
+n(n+1)J^{(0)}_{i+k}\right)\\
 \\
+ b_i\left(J^{(1)}_{i+k+1}+nJ^{(0)}_{i+k+1}\right)
\end{array}
\right\}\tilde\tau_n(t)=0.
$$
 for $\TT(t)$ follows at once
from (6.8) and proposition 4.1, from which (6.9) follows, using the
notation (0.12).\qed

\section{Virasoro constraints with boundary terms }

\setcounter{equation}{0}

As before, consider the matrix integral over symmetric
matrices
\be
\tilde\tau_{2n}(t,E)=\int_{{\cal S}_{2n}(E)}
e^{tr(V(X)+\sum_1^{\iy}t_iX^i)}dX,
\ee
integrated over the space ${\cal S}_{2n}(E)$ of symmetric matrices
with spectrum in $E\subset\BR$, where
\be
E=\mbox{\,disjoint union
$=\displaystyle{\bigcup^r_{i=1}}[c_{2i-1},c_{2i}]$ and $V'(z)=
\displaystyle{\frac{g}{f}=\frac{\sum b_iz^i}{\sum a_iz^i}}.$}
\ee
The purpose of this section is to show that the integral (7.1)
satisfies Virasoro constraints, with an extra-contribution coming
from the boundary of $E$. When $E=\BR$, one recovers the equations
of Theorem 6.2, without the boundary contribution. The method here
hinges on the explicit integral representation of $\TT$ in terms of
an integral over symmetric matrices, whereas the string equation
method does not use that representation, but rather reveals the
rich mathematical structures behind the $\TT$-functions.

 Remember the definition of the Virasoro vectors given as in (0.12)
and the notation $V(x,t)$ of section 5:
\bea
\BJ^{(1)}_k &=&(\BJ^{(1)}_{k,n})_{n\geq
0}=\left(J_k^{(1)}+nJ_k^{(0)}\right)_{n\geq 0},\nonumber\\
\BJ_k^{(2)}&=&(\BJ^{(2)}_{k,n})_{n\geq
0}=\left(J_k^{(2)}+(2n+k+1)J_k^{(1)}+n(n+1)J_k^{(0)}\right)_{n\geq
0}.\nonumber
\eea

Instead, we shall consider a slight generalization of these
Virasoro vectors:
\bea
\BJ^{(1)}_k &=&(\BJ^{(1)}_{k,n})_{n\geq
0}=\left(J_k^{(1)}+nJ_k^{(0)}\right)_{n\geq 0},\nonumber\\
\BJ_k^{(2)}&=&(\BJ^{(2)}_{k,n})_{n\geq
0}=\left(\beta J_k^{(2)}+(2n\beta+(2-\beta)(k+1))J_k^{(1)}+n(\beta
n+2-\beta)J_k^{(0)}\right)_{n\geq 0}.\nonumber
\eea
with
\bea
J_k^{(0)}&:=&\dt_{k0}\nonumber\\ J_k^{(1)}&:=&\frac{\pl}{\pl
t_k}+\frac{1}{\beta}(-k)t_{-k},\hspace{1cm} J_0^{(1)}=0,\nonumber\\
J_k^{(2)} &:=&\sum_{i+j=n}\frac{\pl^2}{\pl t_i\pl
t_j}+\frac{2}{\beta}
\sum_{-i+j=n}it_i\frac{\pl}{\pl t_j}\nonumber
\eea

\vspace{0.2cm}

\begin{theorem} Given the potential $V$ as in (7.2),
the integrals $\tilde\tau_N(t,E)$ satisfy the following Virasoro
constraints
\be
\left(\sum_{i=1}^{2r}c_i^{k+1}f(c_i)\frac{\pl}{\pl c_i}-\sum_{\ell=0}^{\iy}
\left(\frac{a_{\ell}}{2}\BJ_{k+\ell,N}^{(2)}+b_{\ell}
\BJ_{k+\ell+1,N}^{(1)}\right)\right)
\tilde\tau_{N}(t,E)=0
\ee
for all $k\geq -1$, and even $N\geq 0$.
\end{theorem}

\vspace{0.2cm}

\begin{lemma} Given a symmetric $N\times N$ matrix $X$, the following variational
formula holds:
$$\frac{d}{d\vr}d(X+\vr f(X)X^{k+1})e^{tr\,V(X+\vr
f(X)X^{k+1},t)}\Biggl|_{\vr=0}$$
\be=\sum^{\iy}_{\ell=0}\left(\frac{a_{\ell}}{2}\BJ^{(2)}_{k+\ell,N}
+b
_{\ell}\BJ^{(1)}_{k+\ell +1,N}\right)dX\,e^{tr\,V(X,t)}.
\ee
\end{lemma}

\proof At first, note that, in view of (5.4),
\be
dX\,e^{tr\,V(X,t)}=|\Delta_N(x)|^{\beta}e^{\sum^N_{k=1}(V(x_k)+\sum^{\iy}
_{i=1}t_ix^i_k)}dx_1...dx_N\,dO
\ee
and the map $X\mapsto X+\vr f(X)X^{k+1}$ induces a map on the $x_1,...,x_N$:
\be
x_i\mapsto x_i+\vr f(x_i)x_i^{k+1}.
\ee
Also, observe the following two relations for $k\geq 0$:

\medbreak

$\displaystyle{\left(\frac{1}{2}\sum_{{i+j=k}\atop{i,j>0}}\frac{\pl^2}{\pl t_i\pl
t_j}-\frac{N}{2}
\delta_{k,0}\right)e^{tr\,V(X)}
}$
\be
\hspace{3.5cm} =\left(\sum_{{1\leq a<\beta\leq N}\atop{i+j=k,i,j>0}}x^i_{\al}x^j_{\beta}+
\frac{k-1}{2}\sum_{1\leq\al\leq N}x_{\al}^k\right)e^{tr\,V(X)},
\ee
\be
\left(\frac{\pl}{\pl t_k}+N\delta_{k,0}\right)e^{tr\,V(X)}=\left(
\sum_{1\leq\al\leq N}x_{\al}^k\right)e^{tr\,V(X)}.
\ee
So, the point now is to compute the $\vr$-derivative
\be
\frac{d}{d\vr}\left(|\Delta_N(x)|^{\beta}e^{\sum^N_{k=1}(V(x_k)+\sum^{\iy}
_{i=1}t_ix^i_k)}
dx_1...dx_N\right)_{x_i\mapsto x_i+\vr
f(x_i)x_i^{k+1}}\Biggl|_{\vr=0},
\ee
which consists of three contributions:

\medbreak

\noindent\underline{part 1}:

\medbreak

\noindent$\displaystyle{\frac{\pl}{\pl\vr}\left|\Delta(x+\vr
f(x)x^{k+1})\right|^{\beta}\Biggl|_{\vr=0}}$
\bea
&=&\beta|\Delta(x)|^{\beta}\sum_{1\leq\al <\ga\leq
N}\frac{\pl}{\pl\vr}\log\left(|x_{\al}-x_{\ga}+\vr(f(x_{\al})x_{\al}^{k+1}
-f(x_{\ga})x_{\ga}^{k+1})|\right)\Biggl|_{\vr=0}\nonumber\\
&=&\beta|\Delta(x)|^{\beta}\sum_{1\leq\al <\ga\leq
N}\frac{f(x_{\al})x_{\al}^{k+1}-f(x_{\ga})x_{\ga}^{k+1}}
{x_{\al}-x_{\ga}}\nonumber\\
&=&\beta|\Delta(x)|^{\beta}\sum_{\ell=0}^{\iy}a_{\ell}
\sum_{1\leq\al
<\ga\leq N}
\frac{x_{\al}^{k+\ell+1}-x_{\ga}^{k+\ell+1}}{x_{\al}-x_{\ga}}\nonumber\\
&=&\beta|\Delta(x)|^{\beta}\sum_{\ell=0}^{\iy}a_{\ell}\left(\sum_{{i+j=\ell+k}\atop{1\leq\al
<\ga\leq N,i,j>0}}x^i_{\al}x^j_{\ga}+(N-1)\sum_{1\leq\al\leq
N}x_{\al}^{\ell+k}-\frac{N(N-1)}{2}\delta_{\ell+k,0}\right)\nonumber\\
&=&\beta
e^{-tr\,V(X,t)}|\Delta(x)|^{\beta}\sum_{\ell=0}^{\iy}a_{\ell}\Biggl(\frac{1}{2}
\sum_{{i+j=k+\ell}\atop{i,j>0}}\frac{\pl^2}{\pl t_i\pl
t_j}-\frac{N}{2}\delta_{k+\ell,0}\nonumber\\ &
&+\left(N-\frac{k+\ell+1}{2}\right)
\left(\frac{\pl}{\pl t_{k+\ell}}+N\delta_{k+\ell,0}\right)-\frac{N(N-1)}{2}
\delta_{k+\ell,0}\Biggr)e^{tr\,V(X,t)}\nonumber\\
&=&\beta
e^{-tr\,V(X,t)}|\Delta(x)|^{\beta}\sum_{\ell=0}^{\iy}a_{\ell}\nonumber\\
&&
\Biggl(\frac{1}{2}
\sum_{{i+j=k+\ell}\atop{i,j>0}}\frac{\pl^2}{\pl t_i\pl
t_j}+\left(N-\frac{k+\ell+1}{2}\right)
\frac{\pl}{\pl t_{k+\ell}}
+\frac{N(N-1)}{2}
\delta_{k+\ell,0}\Biggr)e^{tr\,V(X,t)}.\nonumber\\
&&
\eea

\medbreak

\noindent\underline{part 2}:

\medbreak

\noindent$\displaystyle{\frac{\pl}{\pl\vr}\prod^N_1d(x_{\al}+\vr
f(x_{\al})x_{\al}^{k+1})\Biggl|_{\vr=0}}$
\bea
&=&\sum^N_1\left(f'(x_{\al})x_{\al}^{k+1}
+(k+1)f(x_{\al})x_{\al}^k\right)\prod^N_1dx_i\nonumber\\
&=&\sum^{\iy}_{\ell=0}(\ell+k+1)a_{\ell}\sum^N_{\al=1}x_{\al}^{k+\ell}
\prod^N_1dx_i\nonumber\\
&=&e^{-tr\,V(X,t)}\sum^{\iy}_{\ell=0}(\ell+k+1)a_{\ell}\left(\frac{\pl}{\pl
t_{k+\ell}}+N\delta_{k+\ell,0}\right)e^{tr\,V(X,t)}
\prod^N_1 ~dx_i,\nonumber\\
&&
\eea

\medbreak

\noindent\underline{part 3}:

\medbreak

\noindent$\displaystyle{\frac{\pl}{\pl\vr}\prod^N_{\al=1}\exp
\left(V \left(x_{\al}+\vr
f(x_{\al})x^{k+1}\right)+\sum^{\iy}_{i=1}t_i\sum^N_{\al=1}\left(
x_{\al}+\vr f(x_{\al})
x_{\al}^{k+1}\right)^i\right)\Biggl|_{\vr=0}}$
\bea
&=&\left(\sum^N_{\al=1}V'(x_{\al})f(x_{\al})x_{\al}^{k+1}+\sum^{\iy}_{i=1}
it_i\sum^N_{\al=1}f(x_{\al})x_{\al}^{i+k}\right)e^{tr\,V(X,t)}\nonumber\\
&=&\left(\sum^{\iy}_{\ell=0}b_{\ell}\sum^N_{\al=1}
x_{\al}^{k+\ell+1}+\sum_{{\ell\geq 0}\atop{i\geq
1}}a_{\ell}it_i\sum_{\al=1}^Nx_{\al}^{i+k+\ell}\right)
e^{tr\,V(X,t)}\nonumber\\
&=&\Biggl(\sum^{\iy}_{\ell=0}b_{\ell}\left(\frac{\pl}{\pl
t_{k+\ell+1}}+N\delta_{k+\ell+1,0}\right)\nonumber\\ & &\quad
+\sum^{\iy}_{\ell=0}a_{\ell}
\sum^{\iy}_{i=1}it_i\left(\frac{\pl}{\pl
t_{i+k+\ell}}+N\delta_{i+k+\ell,0}\right)\Biggr)e^{tr\,V(X,t)}.
\eea
As mentioned, for knowing (7.9), we must add up the three
contributions (7.10), (7.11) and (7.12), resulting in

\medbreak

\noindent$\displaystyle{\frac{\pl}{\pl\vr}d(X+\vr
f(X)X^{k+1})e^{tr\,V(X+\vr f(X)X^{k+1},t)}\Biggl|_{\vr=0}}$
\bea
&=&\Biggl[\sum_{\ell=0}^{\iy}\frac{a_{\ell}}{2} \Biggl\{
\left(\sum_{{i+j=k+\ell}\atop{i,j\geq
1}}\beta\frac{\pl^2}{\pl t_i\pl t_j}+2\sum_{i\geq
1}it_i\frac{\pl}{\pl t_{i+k+\ell}}\right)\nonumber\\ &
&\quad\quad\quad\quad +(2\beta
N+(2-\beta)\ell+k+1)\left(\frac{\pl}{\pl
t_{k+\ell}}+\frac{t_1}{\beta}\delta_{k+\ell,-1}\right)\nonumber\\ &
&+ N(\beta N-\beta+2)
\delta_{k+\ell,0}\Biggr\}+\sum^{\iy}_{\ell=0}b_{\ell}\Biggl(
\frac{\pl}{\pl t_{k+\ell+1}} +N\delta_{k+\ell+1,0}\Biggr)\Biggr]
 dX e^{tr V(X,t)}\nonumber\\
&=&\sum^{\iy}_{\ell=0}\frac{a_{\ell}}{2}\left(J^{(2)}_{k+\ell}
+(2\beta N+(2-\beta)\ell+k+1) J^{(1)}_{k+\ell}+N(\beta N-\beta
+2)\delta_{k+\ell,0}\right)\nonumber\\ & &\quad\quad\quad\quad
+\sum^{\iy}_{\ell=0}b_{\ell}\left(
J^{(1)}_{k+\ell+1}+N\delta_{k+\ell+1,0}\right) dX e^{tr
V(X,t)},\nonumber
\eea
ending the proof of lemma 7.2.\qed

\medbreak

\underline{\sl Proof of Theorem 7.1}: The change of integration variable
\be
X\mapsto Y=X+\vr f(X)X^{k+1}
\ee
in the matrix integral
$$
\int_{{\cal S}_{2n}(E)}e^{Tr\,V(X,t)}dX
$$
leaves the integral invariant, but it induces a change of limits of
integration, given by the inverse of the map (7.14); namely the
$c_i$'s in $E=\displaystyle{\bigcup^r_1}[c_{2i-1},c_{2i}]$, get
mapped as follows
$$
c_i\mapsto c_i-\vr f(c_i)c_i^{k+1}+O(\vr^2).
$$
Therefore, setting
$$E^{\vr}=\displaystyle{\bigcup^r_1}[c_{2i-1}-\vr f(c_{2i-1})
c_{2i-1}^{k+1}+O(\vr^2),c_{2i}-\vr f(c_{2i})c_{2i}^{k+1}
+O(\vr^2)],$$ we find, using (7.4) and the fundamental theorem of
calculus,
\begin{eqnarray*}
0&=&\frac{\pl}{\pl\vr}\int_{(E^{\vr})^{2n}}|\Delta_{2n}(x+\vr
f(x)x^{k+1})|\prod^{2n}_{i=1}e^{V(x_i+\vr f(x_i)x_i^{k+1}
,t)}d(x_i+\vr f(x_i)x_i^{k+1})\\
&=&\left(-\sum^{2r}_{i=1}c_i^{k+1}f(c_i)\frac{\pl}{\pl c_i}
+\sum^{\iy}_{i=0}\left(
\frac{a_i}{2}\BJ^{(2)}_{k+\ell,2n}+b_i\BJ^{(1)}_{k+\ell+1,2n}
\right)\right)\tilde\tau_{2n}(t,E),
\end{eqnarray*}
ending the proof of Theorem 7.1.\qed

\section{Inductive equations for Gaussian
and Laguerre ensembles}

 Consider a time-dependent probability density $$
\frac{e^{V(z)+\sum_1^{\iy} t_i
z^i}~dz}{\int_{\FR}e^{V(z)+\sum_1^{\iy}t_i z^i} dz}$$ on an
interval $\FR \subset \BR$, with $e^{V(z)}$ decaying fast enough to
$0$ at the boundary of $\FR$. The aim of this section is to find an
inductive expression, given the disjoint union $
E=\bigcup^r_1[c_{2i-1},c_{2i}]
\subset\FR $, for the probability
\bea
\mbox{Prob (spectrum }X\in E):=P_{N+2}(t,E)&=&\frac{\int_{{\cal
S}_{N+2}(E)}e^{Tr(V(X)+\sum_1^{\iy}t_iX^i)} dX}{\int_{{\cal
S}_{N+2}(\FR)}e^{Tr(V(X)+\sum_1^{\iy}t_iX^i)} dX}\nonumber\\ &=&
\frac{\TT_{N+2}(t,E)}{\TT_{N+2}(t,\FR)},
\eea
after setting $t=0$, in terms of $P_{N}(0,E)$ and $P_{N-2}(0,E)$.
It hinges on turning on the time $t$ in $P_N(0,E)$ as in (8.1) and
to combine the Virasoro relations for $\TT_{N}(t,E)$, as obtained
in Theorem 7.1,
\be
\left(\sum_{i=1}^{2r}c_i^{k+1}f(c_i)\frac{\pl}{\pl c_i}-\sum_{\ell=0}^{\iy}
\left(\frac{a_{\ell}}{2}\BJ_{k+\ell,N}^{(2)}+b_{\ell}\BJ_{k+\ell+1}^{(1)}\right)\right)
\tilde\tau_{N}(t,E)=0,~~~V'=\frac{g}{f},
\ee
with the non-linear hierarchy of PDE's (0.4) for $k=0$, but
expressed in terms of $\log\TT_N(t,E)$:
\be
\left(\frac{\pl^4}{\pl
t_1^4}+3\frac{\pl^2}{\pl t_2^2}-4
\frac{\pl^2}{\pl t_1\pl t_3}\right)\log\TT_N+6 \left(\frac{\pl^2}{\pl
t_1^2}\log\TT_N\right)^2=12\frac{\TT_{N-2}\TT_{N+2}}{\TT_N^2}.
\ee
Given $E \subset \FR $, define the non-commutative differential
operators
\be
\DR_k=\sum^{2r}_{i=1}c_i^{k+1}\frac{\pl}{\pl c_i},
~\mbox{for}~ k\geq -1.
\ee
We now state:

\begin{theorem} ({\bf Gaussian ensemble})
For even $N$, the probability
\be
P_{N+2}(0,E)=\frac{\int_{{\cal S}_{N+2}(E)}e^{-Tr X^2}}
{\int_{{\cal S}_{N+2}(\BR)}e^{-Tr X^2}}
\ee
is expressed in terms of $P_{N-2}(0,E)$ and the non-commutative
operators $\DR_k$, acting on $P_N(0,E)$, as follows:

\vspace{0.3cm}

\noindent $192 b_N~\displaystyle
{\frac{P_{N+2}(0,E)P_{N-2}(0,E)}{P_N^2(0,E)}}-12N(N-1)=$\hfill
\be
\hspace{0.1cm}(\DR_{-1}^4+8(2N-1)\DR_{-1}^2+12\DR_0^2+
24\DR_0-16\DR_{-1}\DR_1)\log P_N +6(\DR^2_{-1}
\log P_N)^2.
\ee
When the set $E=[c, \iy]$, then  $P_{N+2}(0,E)$ is expressed in
terms of $G(c):=\frac{\pl}{\pl c}
\log
P_N(0,[c,\iy])$, as follows:
\be
192 b_N~\frac{P_{N+2}P_{N-2}}{P_N^2}-12N(N-1)
=G'''+6~G'^2-4\left(\,c^{2}-2\,(2N-1)\right)\,G'+4c~G.
\ee
\end{theorem}

\remark The differential operator appearing on the right hand side
of (8.7) is reminiscent of the Painlev\'e IV equation. In (8.7),
the constant $b_N$ takes on the following form:
$$
b_N^{-1}=\frac{\left(\int_{{\cal S}_{N}(\BR)}e^{-Tr
X^2}\right)^2}{\int_{{\cal S}_{N-2}(\BR)}e^{-Tr X^2}\int_{{\cal
S}_{N+2}(\BR)}e^{-Tr X^2}}
$$

\proof
Setting $V=-z^2$, we have $V'=-2z=:g/f$, with $g=-2z,~f=1$,
 and so:
$$
\begin{array}{llll}
b_0=0,&b_1=-2,&\mbox{all other} \,\,b_i=0& \\
a_0=1,&a_1=0&\mbox{all other} \,\,a_i=0,&.
\end{array}
$$
With these data, the equations (8.2) become for $k\geq
-1$:
$$
\DR_k\TT_N=\frac{1}{2}\left(J_k^{(2)}+(2N+k+1)J_k^{(1)}+
N(N+1)J_k^{(0)} -4 J^{(1)}_{k+2}
\right)\TT_N.
$$
In particular for $k=-1,0,1$, the function $F(t,E):=\log\TT_N(t,E)$
satisfies
\bea
\DR_{-1}F&=&\left(\sum_{i\geq 2}it_i\frac{\pl}{\pl
t_{i-1}}-2\frac{\pl}{\pl t_1}\right)F+Nt_1\nonumber\\
\DR_0F&=&\left(\sum_{i\geq 1}it_i\frac{\pl}{\pl
t_i}-2\frac{\pl}{\pl t_2}\right)F+\frac{N(N+1)}{2}\nonumber\\
\DR_1F&=&\left(\sum_{i\geq 1}it_i\frac{\pl}{\pl
t_{i+1}}-2\frac{\pl}{\pl t_3}+(N+1)\frac{\pl}{\pl
t_1}\right)F.
\eea
Upon taking linear combinations of (8.8), in order to get as
leading term the partials $\pl / \pl t_i$, and upon setting
\be
{\cal B}_1=-\frac{1}{2}\DR_{-1},~~{\cal
B}_2=-\frac{1}{2}\DR_0,~~{\cal
B}_3=-\frac{1}{2}\left(\DR_1+\frac{N+1}{2}\DR_{-1}\right),
\ee
one finds:
\bea
{\cal B}_1 F&=&\left(\frac{\pl}{\pl t_1}-\frac{1}{2}
\sum_{i\geq 2}it_i\frac{\pl}{\pl t_{i-1}}\right)F-\frac{Nt_1}{2}\nonumber\\
{\cal B}_2 F&=&\left(\frac{\pl}{\pl t_2}-\frac{1}{2}
\sum_{i\geq 1}it_i\frac{\pl}{\pl t_i}\right)F-\frac{N(N+1)}{4}\nonumber\\
{\cal B}_3 F&=&\left(\frac{\pl}{\pl t_3}-\frac{1}{2}
\sum_{i\geq 1}it_i\frac{\pl}{\pl
t_{i+1}}-\frac{N+1}{4}\sum_{i\geq 2}it_i\frac{\pl}{\pl
t_{i-1}}\right)F
-\frac{N(N+1)}{4}t_1.\nonumber\\
\eea
These expressions have the precise form (9.1) in the appendix with
$$
\left\{\begin{array}{l}
\ga_{1,-1}=-\frac{1}{2},\ga_{1,0}=\ga_1=0,\dt_1=-\frac{N}{2}\\
\ga_{2,-1}=0,\ga_{2,0}=-1/2,\ga_{2,1}=0,\ga_2=-\frac{N(N+1)}{4},\dt_2=0\\
\ga_{3,-1}=-\frac{N+1}{4},\ga_{3,0}=0,\ga_{3,1}=-\frac{1}{2},\ga_{3,2}=\ga_3=0,
\dt_3=-\frac{N(N+1)}{4}.
\end{array}\right.
$$
Expressing the partial derivatives of $F(t,c):=\log\TT_N(t,c)$ with
respect to the $t_i$'s at $t=0$, in terms of the ${\cal B}_iF$ and
the ${\cal B}_i{\cal B}_j F$, as in (9.3), and setting those into
equation (8.3) leads to:
\be
({\cal B}_1^4+6N{\cal B}_1^2+3{\cal B}_2^2-3{\cal B}_2-4{\cal
B}_1{\cal B}_3)F+6({\cal B}^2_1
F)^2+\frac{3}{4}N(N-1)=12\frac{\TT_{N-2}\TT_{N+2}}{\TT_N^2}.
\ee
Substituting the expressions (8.9) into (8.11) yields:
$$
(\DR_{-1}^4+8(2N-1)\DR_{-1}^2+12\DR_0^2+24\DR_0-16\DR_{-1}\DR_1)F
+6(\DR^2_{-1} F)^2 +12N(N-1)$$
\medbreak

\hfill$=192\displaystyle{\frac{\TT_{N-2}\TT_{N+2}}{\TT_N^2}},$

\noindent which is (8.6). Note that since $F$ appears on the
left hand side, always preceded by a differential operator $\DR_k$
and since $\tau_N(t,\BR)$ is independent of $c$, we may set $F=\log
\frac{\tau_N(0,E)}{\tau_N(0,\BR)}=\log P_N(0,E)$ instead, in the expression above.
 On the right hand side, we substitute $\tilde
 \tau_N(0,E)=P_N(0,E)\tilde \tau_N(0,\BR)$, thus establishing (8.6).
 When the set $E=[c, \iy]$, equation (8.7) follows at once
from (8.6).  \qed

\medbreak

\begin{theorem} ({\bf Laguerre ensemble})
For even $N$ and $c=(c_1,...,c_{2r})$, the probability
\be
P_{N+2}(0,E)= \frac{\int_{{\cal S}_{N+2}(E)}e^{-Tr (X-\al \log X)}
dX}{\int_{{\cal S}_{N+2}(\BR^+)}e^{-Tr (X-\al \log X)} dX},
\ee
is expressed in terms of $P_{N-2}$ and the non-commutative
operators $\DR_k$, acting on $P_N$, as follows:

\vspace{0.3cm}
\noindent $12
b_N~\displaystyle{\frac{P_{N+2}(0,E)P_{N-2}(0,E)}{P_N^2(0,E)}}-\frac{3}{4}N(N-1)(N+2\al)(N+2\al+1)$\hfill

\begin{eqnarray}
&=&\Biggl(\DR_{0}^4-4\DR_{0}^3+\left(3(N+\al)^2-(2\al-1)(2\al+3)\right)
\DR^2_{0}-3(N^2+2\al N-\al)\DR_{0}\nonumber\\
&&\hspace{0.7cm}+3\DR^2_1-3(N+\al)\DR_1+2(N+\al)\DR_{0}\DR_1
+6\DR_2-4\DR_{0}\DR_2\Biggr)
\log P_N\nonumber\\
&&+3(\DR_{0}\log P_N)^2-8(\DR_{0}\log P_N)(\DR_{0}^2\log P_N)+6
(\DR_{0}^2\log P_N)^2.\nonumber\\
\end{eqnarray}

When the set $E=[c, \iy]$, then  $P_{N+2}(0,E)$ is expressed in
terms of $G:=\frac{\pl}{\pl c}
\log
P_N(0,E)$, as follows:

\vspace{0.3cm}
$12b_N~\displaystyle{\frac{P_{N+2}P_{N-2}}{c
P_N^2}}-\frac{3}{4c}N(N-1)(N+2\al)(N+2\al+1)$\hfill

\be=c^{3}\,G'''+2c^2G'' +c \left(3\,N^{2}+2
 \,c\,N+6\,{\rm \al}\,N-\left(c-\,
 {\rm \al}\right)^{2}-4\,{\rm \al}-2\right)~G'$$
 $$+\,\left(c\,N+{\rm \al}\,c-
 {\rm \al}^{2}-{\rm \al}\right)~G
+c\,
 \left(6\,c^{2}\,
 G'^{2}
 +4\,c\,G\,G'+G^{2}
 \right)
 \ee
\end{theorem}

\remark  The right hand side of (8.14) is the expression appearing in
Painlev\'e V.
 In (8.13) and (8.14), the constant $b_N$ takes on the following form:
$$
b_N^{-1}=\frac{\left(\int_{{\cal S}_{N}(\BR^+)}e^{-Tr (X-\al \log
X)}\right)^2}{\int_{{\cal S}_{N-2}(\BR^+)}e^{-Tr (X-\al \log
X)}\int_{{\cal S}_{N+2}(\BR^+)}e^{-Tr (X-\al \log X)}}
$$

\proof
Setting $e^{V(z)}=z^{\al}e^{-z}$, we have $V'=\frac{\al-z}{z}$, and
so $g=\al -z$ and $f=z$, with
$$
\begin{array}{llll}
b_0=\al,&b_1=-1&,\mbox{ all other} \,\,b_i=0& \\
a_0=0,&a_1=1&,\mbox{ all other} \,\,a_i=0,&.
\end{array}
$$
The equations (8.2) become for $k\geq
-1$:

\medbreak

\noindent $
\DR_{k+1}\TT_N$\hfill
$$=\left(\frac{1}{2}J^{(2)}_{k+1}+\frac{1}{2}(2N+k+2\al+2)J_{k+1}^{(1)}+
\frac{1}{2}N(N+2\al+1)J^{(0)}_{k+1}-J^{(1)}_{k+2}\right)\TT_n,
$$
and so for $k=-1,0,1$, the function $F:=\log\TT_N$ satisfies the
following equations:
\bea
\DR_{0}F&=&\left(-\frac{\pl}{\pl t_1}+\sum_{i\geq 1}it_i\frac{\pl}{\pl
t_i}\right)F+\frac{1}{2}N(N+2\al+1)\nonumber\\
\DR_1F&=&\left(-\frac{\pl}{\pl t_2}+(N+\al+1)\frac{\pl}{\pl t_1}+
\sum_{i\geq 1}it_i\frac{\pl}{\pl t_{i+1}}\right)F\nonumber\\
\DR_2F&=&\left(-\frac{\pl}{\pl t_3}+\left(N+\al+\frac{3}{2}\right)\frac{\pl}{\pl
t_2}+\sum_{i\geq 1}it_i\frac{\pl}{\pl
t_{i+2}}+\frac{1}{2}\frac{\pl^2}{\pl t_1^2}\right)F\nonumber\\
&&\hspace{6cm}+\frac{1}{2}\left(\frac{\pl}{\pl t_1}\log
F\right)^2.\nonumber
\eea
Upon taking appropriate linear combinations of the equations above
and setting
\bea
\CB_1&=&-\DR_{0}\nonumber\\
\CB_2&=&-\DR_1-(N+\al+1)\DR_{0}\nonumber\\
\CB_3&=&-\DR_2-\left(N+\al+\frac{3}{2}\right)\DR_1-\left(N+\al+\frac{3}{2}\right)
(N+\al+1)\DR_{0},
\eea
one finds
\bea
\CB_1F&=&\left(\frac{\pl}{\pl t_1}-\sum_{i\geq 1} it_i\frac{\pl}{\pl t_i}\right)F-\frac{N}{2}
(N+2\al+1)\nonumber\\
\CB_2F&=&\left(\frac{\pl}{\pl t_2}-
(N+\al+1)\sum_{i\geq 1} it_i\frac{\pl}{\pl t_i}-\sum_{i\geq 1}
it_i\frac{\pl}{\pl t_{i+1}}\right)F\nonumber\\ &
&\hspace{4cm}-\frac{N}{2}(N+\al+1)(N+2\al+1)\nonumber\\
\CB_3F&=&\Biggl(\frac{\pl}{\pl t_3}-\sum_{i\geq 1}it_i\frac{\pl}{\pl
t_{i+2}}-\left(N+\al+\frac{3}{2}\right)(N+\al+1)\sum_{i\geq
1}it_i\frac{\pl}{\pl t_i}\nonumber\\ &
&\hspace{4cm}-\left(N+\al+\frac{3}{2}\right)\sum_{i\geq
1}it_i\frac{\pl}{\pl t_{i+1}}\Biggr)F\nonumber\\
& &
-\frac{1}{2}\left(\frac{\pl^2F}{\pl t^2_1}+
\left(\frac{\pl F}{\pl t_1}\right)^2\right)-\frac{N}{2}(N+1+\al)(N+1+2\al)\left(N+
\frac{3}{2}+\al\right).\nonumber
\eea
Then refering to the appendix, one sets $\dt_1=\dt_2=\dt_3=0$, and
$$
\begin{array}{l}
\ga_{1,-1}=0,\ga_{1,0}=-1,\ga_1=-\frac{N}{2}(N+1+2\al),\\
\ga_{2,-1}=0,\ga_{2,0}=-(N+\al+1),\ga_{2,1}=-1,\ga_2=-\frac{N}{2}(N+\al+1)
(N+2\al+1),\\
\ga_{3,-1}=0,\ga_{3,0}=-(N+\al+1)(N+\al+\frac{3}{2}),\ga_{3,1}=-(N+\al+\frac{3}{2}),
\ga_{3,2}=-1,\\
\hspace{5cm}\ga_3=-\frac{N}{2}(N+\al+\frac{3}{2})(N+\al+1)(N+2\al+1),
\end{array}
$$
to find
\begin{eqnarray*}
&&\Biggl(\CB_1^4+4\CB_1^3+2(2N^2+2(2\al+1)N+3)\CB_1^2
+3(N^2+(2\al+1)N+1)\CB_1\\
&&\hspace{4cm}+3\CB^2_2+6(N+\al+1)\CB_2-6\CB_3-4\CB_1\CB_3\Biggr)\log\TT_N\\
&&+3(\CB_1\log\TT_N)^2+6(\CB_1^2\log\tau_N)^2+
8(\CB_1\log\TT_N)(\CB^2_1\log\TT_N)\\
&=&-\frac{3}{4}N(N-1)(N+2\al)(N+2\al+1)~+~12~\frac{\TT_{N-2}\TT_{N+2}}{\TT^2_N}.
\end{eqnarray*}
Expressing the $\CB_i$ in terms of the $\DR_i$ according to (8.15),
one finds:\footnote{$[\DR_{0},\DR_1]=\DR_1$,
$[\DR_{0},\DR_2]=2\DR_2$}
$$
\Biggl(\DR_{0}^4-4\DR_{0}^3+\left(3(N+\al)^2-(2\al-1)(2\al+3)\right)
\DR^2_{0}-3(N^2+2\al N-\al)\DR_{0}$$
\begin{eqnarray*}
&&\hspace{1cm}+3\DR^2_1-3(N+\al)\DR_1+2(N+\al)\DR_{0}\DR_1+6\DR_2-4\DR_{0}\DR_2\Biggr)
\log\TT_N\\
&&+3(\DR_{0}\log\TT_N)^2-8(\DR_{0}\log\TT_N)(\DR_{0}^2\log\TT_N)+6
(\DR_{0}^2\log\TT_N)^2\\
&=&-\frac{3}{4}N(N-1)(N+2\al)(N+2\al+1)~+~12\frac{\TT_{N-2}\TT_{N+2}}{\TT_N^2}.
,\end{eqnarray*} which amounts to equation (8.13), upon using the
same kind of argument as in the Gaussian ensemble. Specializing to
the case of a single semi-infinite interval, yields equation
(8.14).\qed

\bigbreak


\section{ Appendix}

\medbreak

Given first order operators $\CB_1,\CB_2,\CB_3$ in $c=(c_1,...,c_{2r})\in\BR^{2r}$
and a function $F(t,c)$, with $t\in\BC^{\iy}$. Let $F$ satisfy the following
partial differential equations in $t$ and $c$:
\be
\CB_kF=\frac{\pl F}{\pl t_k}+\sum_{-1\leq
j<k}\ga_{kj}V_j(F)+\ga_k+\delta_kt_1,\quad k=1,2,3,
\ee
with
\be
V_j(F)=\sum_{i,i+j\geq 1}it_i\frac{\pl F}{\pl
t_{i+j}}+\frac{1}{2}\delta_{2,j}\left(\frac{\pl^2 F}{\pl
t_1^2}+\left(\frac{\pl F}{\pl t_1}\right)^2\right),\quad -1\leq
j\leq 2.
\ee
Then, at $t=0$, one computes

\begin{eqnarray*}
\frac{\pl F}{\pl t_1}\Biggl|_{t=0}&=&\CB_1F-\ga_1\nonumber\\
\frac{\pl^2 F}{\pl
t^2_1}\Biggl|_{t=0}&=&\left(\CB_1^2-\ga_{10}\CB_1\right)F+\ga_{10}\ga_1-\dt_1\nonumber\\
\frac{\pl^3 F}{\pl
t^3_1}\Biggl|_{t=0}&=&\left(\CB_1^3-3\ga_{10}\CB_1^2+
2\ga_{10}^2\CB_1\right)F+2\ga_{10} (\dt_1-\ga_1\ga_{10})\nonumber\\
\frac{\pl^4 F}{\pl
t^4_1}\Biggl|_{t=0}&=&\left(\CB_1^4-6\ga_{10}\CB_1^3+11\ga_{10}^2\CB_1^2-6\ga_{10}^3
\CB_1\right)F-6\ga_{10}^2(\dt_1-\ga_1\ga_{10})\nonumber\\
\frac{\pl F}{\pl t_2}\Biggl|_{t=0}&=&\CB_2F-\ga_2\nonumber\\
\frac{\pl^2 F}{\pl
t^2_2}\Biggl|_{t=0}&=&\biggl(\CB_2^2-2\ga_{20}\CB_2+\ga_{21}
\ga_{32}\CB^2_1-((2\ga_1 +\ga_{10})\ga_{21}\ga_{32}+2\ga_{2,-1})
\CB_1\\
&&-2\ga_{21}\CB_3\biggr)F +\ga_{21}\ga_{32}(\CB_1F)^2\\ &&
+\ga_{21}\ga_{32}(\ga^2_1+\ga_{10}
\ga_1-\dt_1)+2(\ga_{21}\ga_3+\ga_{20}\ga_2+\ga_1\ga_{2,-1})\nonumber\\
\frac{\pl
F}{\pl
t_3}\Biggl|_{t=0}&=&\left(\CB_3-\frac{\ga_{32}}{2}\CB_1^2+\frac{\ga_{32}}{2}
(2\ga_1+\ga_{10})\CB_1\right)F-\frac{\ga_{32}}{2}(\CB_1F)^2\nonumber\\
&
&+\frac{\ga_{32}}{2}(\dt_1-\ga_1\ga_{10}-\ga^2_1)-\ga_3\nonumber\\
\frac{\pl^2
F}{\pl t_1\pl
t_3}\Biggl|_{t=0}&=&\biggl(\CB_1\CB_3-\frac{\ga_{32}}{2}\CB_1^3
+\ga_{32}( \ga_1+2\ga_{10})\CB_1^2-\frac{3}{2}
\ga_{10}\ga_{32}(2\ga_1+\ga_{10})\CB_1\nonumber\\
&
&-3\ga_{1,-1}\CB_2-3\ga_{10}\CB_3\biggr)F+\frac{3}{2}\ga_{10}\ga_{32}(\CB_1F)^2
-\ga_{32}(\CB_1F)(\CB_1^2F)\nonumber\\
& &+
\frac{3}{2}(2\ga_{10}\ga_3+\ga_{32}\ga_{10}(\ga^2_1+\ga_{10}\ga_1-\dt_1)+2\ga_{1,-1}
\ga_2).\nonumber
\end{eqnarray*}
\be
\ee
Indeed, the method consists of expressing $\displaystyle{\frac{\pl
F}{\pl t_k}\Biggl|_{t=0}}$ in terms of
$\displaystyle{\CB_kf\Biggl|_{t=0}}$, using (9.1). Second
derivatives are obtained by acting on $\CB_kF$ with $\CB_{\ell}$,
by noting that $\CB_{\ell}$ commutes with all $t$-derivatives, by
using the equation for $\CB_{\ell}F$; and by setting in the end
$t=0$:
\bea
\CB_{\ell}\CB_kF&=&\CB_{\ell}\frac{\pl F}{\pl t_k}
+\sum_{-1\leq j<k}\ga_{kj}\CB_{\ell}(V_j(F))\nonumber\\
&=&\left(\frac{\pl }{\pl t_k} +\sum_{-1\leq
j<k}\ga_{kj}V_j\right)\CB_{\ell}(F),\quad\mbox{provided $V_j(F)$
does not}\nonumber\\ & &\hspace{5cm}\mbox{contain non-linear
terms}\nonumber\\ &=&\left(\frac{\pl }{\pl t_k} +\sum_{-1\leq
j<k}\ga_{kj}V_j\right)\left(\frac{\pl F}{\pl t_{\ell}}
+\sum_{-1\leq j<\ell}\ga_{\ell
j}V_j(F)+\dt_{\ell}t_1\right)\nonumber\\ &=&\frac{\pl^2F}{\pl
t_k\pl t_{\ell}}+\mbox{\,lower-weight terms.}\nonumber
\eea
When the non-linear term is present, it is taken care as follows:
\begin{eqnarray*}
\CB_{\ell}\left( \frac{\pl F}{\pl t_1}   \right)^2
&=&2\frac{\pl F}{\pl t_1} \CB_{\ell}  \frac{\pl F}{\pl t_1}\\ &=&
2\frac{\pl F}{\pl t_1}  \frac{\pl }{\pl t_1}       \CB_{\ell}F\\
&=&2\frac{\pl F}{\pl t_1} \frac{\pl }{\pl t_1}
\left(  \frac{\pl F}{\pl t_{\ell}}+\sum_{-1\leq
j<{\ell}}\ga_{{\ell}j}V_j(F)+\ga_{\ell}+\delta_{\ell}t_1 \right)\\
&=&2\frac{\pl F}{\pl t_1}
\left(  \frac{\pl^2 F}{\pl t_{\ell}\pl t_1}+\sum_{-1\leq
j<{\ell}}\ga_{{\ell}j}V_j\left(\frac{\pl F}{\pl
t_1}\right)+\delta_{\ell}
\right);
\end{eqnarray*}
higher derivatives are obtained in the same way.

\end{document}